\documentclass[tightenlines,aps,prc,eqsecnum,epsfig,floats,amsmath,amssymb]{revtex4}
\usepackage{epsfig,graphics,color}
\usepackage{graphicx}% Include figure files
\usepackage{dcolumn}% Align table columns on decimal point
\usepackage{bm}% bold math
\newcommand \ie {{\it i.e.}}

\newcommand \kd  {\delta}

\newcommand \vn {\vec{n}}

\newcommand \lc {\langle}
\newcommand \rc {\rangle}
\newcommand \ra {\rightarrow}

\newcommand \nt {\noindent}

\newcommand \bvec{\left( \begin{array}{c} }
\newcommand \evec{\end{array} \right)}

\newcommand \eg {{\it e.g.}}  
\newcommand \etc {{\it etc.}}  
\newcommand \bea{\begin{eqnarray} }
\newcommand \eea{\end{eqnarray} }
\newcommand \nn {\nonumber}
\newcommand {\be} {\begin{equation}}
\newcommand {\ee} {\end{equation}}

\begin{document}
\preprint{LBNL-52685}

\title{The chemical equilibration volume: measuring the degree of thermalization}

\author{A. Majumder and V. Koch}

\affiliation{Nuclear Science Division, 
Lawrence Berkeley National Laboratory,
1 Cyclotron road, Berkeley, CA 94720.}

\date{ \today}

%\maketitle

\begin{abstract}
We address the issue of the degree of equilibrium achieved in a high 
energy heavy-ion collision. 
%It has been proposed that kinetic equilibrium 
%will be achieved early and will be followed by chemical equilibration.
Specifically, we explore the consequences of incomplete 
strangeness chemical equilibrium. 
This is achieved over a volume $V$ of the 
order of the strangeness correlation length and is assumed to be smaller than 
the freeze-out volume $V_f$.  
%chemical equilibrium is assumed to have been established over a 
%volume $V$ much smaller than the freeze-out volume $V_f$. 
%The main influence of 
%such a situation is shown to be on the probability distributions 
%of strange particles, which are not brought in by the `in state'. 
%Strict strangeness conservation is implemented over this small volume 
%via the canonical ensemble.   
Probability distributions of strange hadrons emanating from 
the system are computed for varying sizes of $V$ and simple experimental 
observables based on these are proposed. Measurements of such observables
may be used to estimate $V$ and as a result the degree of 
strangeness chemical equilibration achieved. This sets a lower bound on the 
degree of kinetic equilibrium. We also point out that a determination of 
two-body correlations or second moments of the distributions are not sufficient 
for this estimation.
%It is probable that complete chemical or even thermal equilibrium may not have
%been reached over the entire volume of the fireball. 
%Such a scenario wields minimal 
%influence on the abundances and momentum distributions of produced hadrons, 
%which appear 
%thermal due to the maximal filling of phase space. 
%As no strange hadrons 
%constitute the `in state', the production of such hadrons serves as an 
%indicator of the degree of chemical equilibration, or the the size of 
%the volume over which chemical equilibration has been achieved. This serves 
%as a lower bound on the thermal equilibration volume. We propose a series 
%of mits and measurements 
%which may be used to estimate this volume. The occurrence of a 
%large chemical equilibration volume does not merely establish the presence 
%of widespread thermalization in a heavy-ion collision, but points 
%towards the presence of a quark gluon plasma earlier in the collision.   
\end{abstract}


\maketitle

%%%%%%%%%%%%%%%%%%%%%%%%%%%%%%
%%%%%%%%%%%%%%%%%%%%%%%%%%%%%%
%%%%%%%%%%%%%%%%%%%%%%%%%%%%%%

\section{INTRODUCTION}

%%%%%%%%%%%%%%%%%%%%%%%%%%%%%%
%%%%%%%%%%%%%%%%%%%%%%%%%%%%%%
%%%%%%%%%%%%%%%%%%%%%%%%%%%%%%

Experiments are presently underway at the relativistic 
heavy-ion collider (RHIC) at BNL and the SPS at CERN 
to collide large nuclei 
at ultra-relativistic energies. The aim is to create energy 
densities high enough for the production of a
state of deconfined quarks and gluons: the quark gluon plasma (QGP). 
Inherent in this statement is the assumption that, in such a collision, 
we have 
created a system close to kinetic and possibly chemical equilibrium.
The QGP is rather short lived and soon hadronizes, leading one 
to observe a plethora of mesons and baryons at the detectors.
One has to surmise by indirect means as to whether a QGP or at least a 
thermalized system was created in the history of a given collision. 
On confronting experimental results, two possibilities
of what may have occurred in such collisions emerge. 

Even though most 
energetic partons are expected to undergo small angle (soft) 
scattering, it is possible that a sizable number of 
partons undergo large angle (hard) scattering followed 
by multiple rescattering in the central region. 
This may lead to the appearance of sufficient energy densities 
in the central region of 
such collisions that quarks and gluons that are usually 
bound in nuclei become deconfined over a region much larger 
than the size of a nucleon. One obtains a thermalized 
Quark Gluon Plasma (QGP) \cite{col75,shu80}. Such a 
state will undoubtably be
short lived and is expected to quickly expand and hadronize into a 
hot plasma of mesons and baryons. The system will continue to 
expand and cool, finally interactions freeze out and 
the hadrons free stream to the detectors. 

The above paragraph represents a rather optimistic possibility.
The other extreme is to view the above collision as a superposition 
of a series of nucleon nucleon collisions. Each such collision 
produces jets of hard hadrons and some soft hadrons as is the 
case in a proton-proton (or proton-antiproton) collision. 
The hadrons mostly emerge without interaction from the central region.
There is no multiple partonic rescattering, and as a result 
no widespread thermalization entails and no quark gluon plasma is formed.
Another equivalent statement, would be that that the quarks and gluons 
that eventually lead to the formation of the hadrons are not deconfined 
over a region much larger than the volume of a nucleon.   

It is reasonable to assume that actual 
heavy-ion experiments lie somewhere in between these two extremes.
To obtain complete thermalization one needs an infinite amount of 
time, not available in a heavy-ion collision. On the other hand, it is
rather unlikely that every hadron produced in each of the nucleon 
nucleon collisions 
will emerge untouched from a region with a density many times that 
of normal nuclear matter. Undoubtedly, there is some thermalization: 
we picture a scenario where an impinging nucleon has a probability 
distribution to undergo $n$ hard or semi-hard scatterings; each of 
these $n$ points becomes a source of particles (see \eg, Ref. \cite{hwa90}). There is also 
a probability distribution of the number of particles produced at these
$n$ points, greater particle production leads to a greater probability 
of final state multiple scattering. Multiple scattering leads to the onset of 
kinetic followed by chemical equilibrium. As a result, one obtains 
multiple small domains in the larger volume, each, very 
close to being  in chemical and kinetic equilibrium. 
The size of the chemically equilibrated volume is 
assumed to be smaller than the kinetically equilibrated volume. 
The particles in these domains can undergo further 
rescattering with particles in surrounding domains leading 
to an increase in the domain size. 
If very little rescattering occurs then the size of these 
domains is closer to the size of a nucleon.
Extensive rescattering leads to domain sizes of the order of the 
size of the system. We will concentrate on the size of a 
chemically equilibrated domain. Whether the size of such domains is  
of the order of the system size or the size of a nucleon will be 
the primary question addressed in this article.

The primary observables in such reactions are the populations of 
the various species of hadrons \cite{bra99,bra01,raf01,zsc02} 
and their momentum distributions \cite{bro01,bro02,fri03}.
Interestingly, the broad characteristics of these observables 
fail to distinguish between the above two scenarios \cite{koc02}. 
The reason behind this strife is the observation that the 
abundances of various hadrons produced in proton-proton, or even $e^+ \, e^-$ 
collisions may be well reproduced by thermal 
models \cite{bec97}. To distinguish between these two possibilities 
one needs to look at finer features of the data (see \eg,\cite{jeo00,ble00}). 
One such 
feature is to look at the momentum spectra of the various 
hadrons, especially at high transverse momenta \cite{adl02}.
Yet another feature is the abundances of hadrons which carry  
a conserved quantum number \cite{ada02}. In this article we will be
concerned with the latter aspect.

There are many exactly conserved and almost exactly conserved 
quantum numbers in a 
heavy-ion collision. For the moment we will specifically concentrate on the 
mid-rapidity or central region (the full size of the region must be such that 
at freeze-out, the boundaries of the system are not beyond causal contact). 
We will discuss the effects on 4$\pi$ observables later.  
Net baryon number is exactly conserved. 
As the fireball lives for a time on the order of the strong interaction scale,
weak interactions and hence flavor changing currents are considered absent. Thus 
net strangeness, and charm are almost exactly conserved. The application 
of thermal models to particle production involving conserved quantum numbers
requires the use of the canonical ensemble for these numbers. 
This was first pointed out in proton-proton collisions in Ref. \cite{hag68}.

If the number of particles are large 
(or we are dealing with a rather large system), 
one approaches the grand canonical 
limit, where one may simply introduce a Lagrange multiplier for the  
conserved quantity and 
evaluate the much simpler grand canonical partition function.
One may then fine tune the Lagrange multiplier to 
obtain the correct value of the conserved number on the average.   
If the number of particles are small, (or we are dealing with a smaller
system), then canonical methods yield results in marked 
distinction from grand canonical estimates \cite{cle91,ham00}.

Using such models one may obtain a better fit to the data on 
particle abundances, both for those influenced by conservation laws and 
those free of such constraints. The constraint that we will 
be concerned with will be strangeness conservation. As the 
partons(hadrons) rescatter they will produce strange partons(hadrons).
This is the onset of chemical equilibration. By the time freeze-out 
occurs, chemical equilibrium has taken place over a volume 
possibly smaller than the system size, this will be referred 
to as the domain or partition volume. The radius of such a domain will, in principle, be governed by the 
strangeness diffusion constant.  We are assuming the full 
system to be made up of multiple such domains. 
%Kinetic equilibrium, which occurs faster, is 
%prevalent over a larger volume. 
%Hence the chemically equilibrated volume 
%is smaller than the kinetically equilibrated volume, 
%and provides a 
%lower bound. 
Within each domain we will enforce exact strangeness conservation 
by calculating strange particle yields using the canonical ensemble.
Particles not subject to this constraint will have their yields 
calculated via a grand canonical ensemble with a baryon chemical potential $\mu$.
The chemical potential will be assumed to be a constant throughout the system; in 
this effort we will set it to zero.
It should be pointed out that in a grand canonical 
calculation, the number of particles is strictly proportional to the volume. This implies  
that the yields of particles not constrained by strangeness is unaffected by the 
partitioning of the system into domains. The pion yield, for example, is directly 
proportional to the total freeze-out volume. 
In the following we will propose measurable observables
that may be used to discern the domain volume (or the strangeness chemical equilibration volume). 

Once again, we point out that a small domain 
volume does not necessarily indicate a small kinetically equilibrated volume. 
The entire system may have reached kinetic equilibrium and frozen out 
before having reached strangeness chemical equilibrium. The strange 
quark has a current mass more than ten times that of the 
up or down quark, while the kaons are about four times as heavy as 
the pions. Hence, if it turned out that the domain volume 
is of the order of the size of the system, then it would 
point towards the presence of an efficient strangeness equilibration 
mechanism in the history of the collision. The leading candidate
for such a mechanism is the quark gluon plasma.

In the following, we begin by outlining 
the calculation of the partition function
for the strange sector for a single 
domain in Sect. II. Here we will introduce the recursion 
relation technique which greatly 
simplifies the numerical calculation compared to the previous 
methods involving Bessel function 
techniques. In Sect. III we compute the probability 
to obtain $A$ distinguishable 
particles from the full freeze-out volume. Here a second 
recursion relation is introduced. 
With the calculation of the probability distributions, various 
statistical quantities may be 
computed as a function of the number of domains. This will be performed and 
the results discussed in Sect. IV. 
The reader not interested in the 
calculational details may 
skip directly to this section. 
Here, we will propose measurable 
observables that will allow 
one to narrow down the range of 
domain volumes. 
Our conclusions are summarized in Sect. V.

%%%%%%%%%%%%%%%%%%%%%%%%%%%%%%%%%%%%%%%%%%%%%%%%%%%
%%%%%%%%%%%%%%%%%%%%%%%%%%%%%%%%%%%%%%%%%%%%%%%%%%%
%%%%%%%%%%%%%%%%%%%%%%%%%%%%%%%%%%%%%%%%%%%%%%%%%%%

\section{The partition function of a domain.}

%%%%%%%%%%%%%%%%%%%%%%%%%%%%%%%%%%%%%%%%%%%%%%%%%%%
%%%%%%%%%%%%%%%%%%%%%%%%%%%%%%%%%%%%%%%%%%%%%%%%%%%
%%%%%%%%%%%%%%%%%%%%%%%%%%%%%%%%%%%%%%%%%%%%%%%%%%%

Let us assume that freeze-out occurs in a large volume of 
size $V_f$. If the system is not chemically equilibrated over this
volume, but over a much smaller volume $V$, we divide up the system
into $p$ partitions or domains, each with a volume $V = V_f/p$. 
In each domain we assume, we have full chemical and thermal 
equilibrium. Strangeness is exactly conserved in a domain, hence
the partition function for net strangeness equal to $N$ is:

\bea
Z(N) &=& Z_0 \sum_{n_1 = 0}^{\infty} \frac{S_1^{n_1}}{n_1 !}
\sum_{n_{-1} = 0}^{\infty} \frac{S_{-1}^{n_{-1}}}{n_{-1} !}
\sum_{n_2 = 0}^{\infty} \frac{S_2^{n_2}}{n_2 !} 
\sum_{n_{-2} = 0}^{\infty} \frac{S_{-2}^{n_{-2}}}{n_{-2} !}
\sum_{n_3 = 0}^{\infty} \frac{S_3^{n_3}}{n_3 !} 
\sum_{n_{-3} = 0}^{\infty} \frac{S_{-3}^{n_{-3}}}{n_{-3} !} \nn \\
& & \times \kd_{n_1 - n_{-1} + 2 n_2 - 2 n_{-2} + 3 n_3 - 3 n_{-3} , N}. 
\label{gnrl_part}
\eea

\nt
In the above equation, $Z_0$ is the grand canonical partition
function for non-strange particles:

\bea
Z_0 = \prod_{i} \exp \left( \int \frac{d^3 r_i d^3 p_i}
{ (2 \pi \hbar)^3 }  
e^{-\beta (E_i - \mu_i)} \right),
\eea

\nt
where, $i$ runs over all the different kinds of non-strange 
particles and anti-particles. The $\mu_i$'s represent the
respective chemical potentials. In Eq. (\ref{gnrl_part}) 
The $S_k$'s represent the single particle partition functions
for particles with strangeness $k$. For instance for strangeness = -1,

\bea
S_1 = S(K^-) + S(\bar{K}^0) + S(K^*) + S(K_1) + S(\Lambda)  + S(\Sigma) + ... \label{sk}
\eea

\nt
In Eq. (\ref{gnrl_part}), the Kroneker delta at the 
end ensures that the total strangeness is held fixed at $N$
(an integer value). The $n!$ terms in the denominators 
serve the dual purpose of symmetrization and the removal of 
double counting. In this paper we will deal with temperatures
where quantum statistics will not play a role; in such a setting
the $n!$ terms are introduced to avoid Gibbs' paradox \cite{rei85}

On calculating the partition function, one may evaluate its 
various derivatives to obtain various quantities of interest. 
The usual method used up to now is to take the Fourier transform
of the Kroneker delta function, which allows one to express the 
strange partition function $Z(N)/Z_0$ in a closed 
form \cite{cle91}:

\bea
Z_S(N) &=& \frac{Z(N)}{Z_0} 
= \frac{1}{2\pi} \int_{-\pi}^{\pi} d \phi 
\exp \left( S_0 + S_1 e^{i \phi} + S_{-1} e^{-i \phi} 
+ S_2 e^{2 i \phi} + S_{-2} e^{-2i \phi} + S_3 e^{3 i \phi}
S_{-3} e^{-3 i \phi} \right).
\eea

\nt  
The solution of this equation however involves the use of 
series of sums of modified Bessel functions 
(see Eq. (24) of Ref. \cite{cle91}). 
This turns out to be rather complicated to handle, both in 
analytic form and in numerical computation. If one is not
interested in an analytic expression but rather in the 
numerical results, a simple and elegant means of 
calculating in the canonical ensemble exists. This involves the
use of recursion relation techniques first introduced by Mekjian 
for intermediate energy heavy-ion reactions \cite{cha95,das98,das99}.
The use of such techniques to solve the problem of exact 
strangeness conservation will be introduced in the following
subsection. The method of recursion relations is very general and 
may be used to calculate a variety of problems with 
exact conservation conditions \eg, baryon number, charm, charge.
The recursion relations can also be used to calculate partition 
functions with multiple conserved charges \cite{maj03}.  
A variant of this method will be used later to calculate 
strangeness production from all the domains together \ie, for the 
full system.

%%%%%%%%%%%%%%%%%%%%%%%%%%%%%%%%%%%%%%%%%%%%%%%%%%%
%%%%%%%%%%%%%%%%%%%%%%%%%%%%%%%%%%%%%%%%%%%%%%%%%%%
%%%%%%%%%%%%%%%%%%%%%%%%%%%%%%%%%%%%%%%%%%%%%%%%%%%

\subsection{The Recursion Relation Technique} 

%%%%%%%%%%%%%%%%%%%%%%%%%%%%%%%%%%%%%%%%%%%%%%%%%%%
%%%%%%%%%%%%%%%%%%%%%%%%%%%%%%%%%%%%%%%%%%%%%%%%%%%
%%%%%%%%%%%%%%%%%%%%%%%%%%%%%%%%%%%%%%%%%%%%%%%%%%%

We begin by expanding Eq. (\ref{gnrl_part}) for the simple and 
realistic case of $N=0$. Once again, we are using 
strangeness to illustrate the derivation of the 
recursion relations, however any conserved charge may be 
used. 
Making explicit use of  the Kroneker delta function, we obtain, 

\bea
Z_S = \frac{Z}{Z_0} &=& 1 + S_1 S_{-1} + 
\frac{ S_1^2 }{2!} \frac{ S_{-1}^2 }{2!}  +  \frac{ S_1^2 }{2!} S_{-2} 
+ S_2 \frac{ S_{-1}^2 }{2!} + ... \nn \\  
&=& 1 + S_1 S_{-1}  +  \left(  \frac{ S_1^2 }{2!} + 
S_2 \right) \left( \frac{ S_{-1}^2 }{2!}  + S_{-2} \right)  + ...
\eea 

\nt
Where we have rewritten $Z(0)$ as simply $Z$. 
In the above series, we identify the various 
terms in the brackets as, 

\bea 
Z_S = \frac{Z}{Z_0} &=&  1 + Z_1 Z_{-1} +  Z_2 Z_{-2} + ... + Z_{n} Z_{-n} + ...  
\label{Z_series}
\eea

\nt where $Z_n$ is the partition function of a box containing 
only strange particles, with net strangeness of the box equal to $n$. 
The term $Z_{-n}$ represents a similar 
system with only antistrange 
particles with net strangeness $-n$.  One immediately 
notes that each term in the sum 
has net strangeness zero. The general term $Z_n$ may 
be expressed formally as 

\bea
Z_n = \sum_{n_1,n_2,... | \sum k n_k = n}  \prod_k \frac{S_k^{n_k}}{n_k!} .
\label{z_n}
\eea

\nt We will eventually set 
$S_k$ to 0 for all $k$ greater than 3 \ie, there are no baryons with 
strangeness
greater than 3; presently we keep them in for generality. 
The constraint on the 
numbers $n_k$, as the subscript on the sum indicates, is such that the 
total strangeness sums to $n$. This implies, that, with these value of $n_k$

\bea
\frac{1}{n} \sum_k k n_k = 1.
\eea
  
\nt We now insert unity, as obtained in the above equation, into 
Eq. (\ref{z_n}), to obtain,

\bea
Z_n &=& \sum_{n_1,n_2,... | \sum k n_k = n}  \frac{1}{n} \sum_k  k n_k \,\, \left[\prod_j 
\frac{S_j^{n_j}}{n_j!} \right] \nn \\
&=& \sum_{n_1,n_2,... | \sum k n_k = n} \frac{1}{n} \sum_k  k n_k
\frac{S_1^{n_1}}{n_1!} 
... \frac{S_k^{n_k}}{n_k!} ... ,
\eea

\nt where we have expanded out the product. We now cancel 
the appearance of $n_k$ in numerator and denominator, and may 
reverse the order of the sums to obtain,

\bea
Z_n &=& \sum_{n_1,n_2,... | \sum k n_k = n} \frac{1}{n} \sum_k  k 
\frac{S_1^{n_1}}{n_1!} 
... \frac{S_k^{n_k}}{(n_k - 1)!} ... \nn \\
&=& \frac{1}{n} \sum_k S_k \sum_{n_1,n_2,... | \sum k n_k = n-k} 
\frac{S_1^{n_1}}{n_1!} ... \frac{S_k^{n_k}}{(n_k)!}   \nn \\
&=& \frac{1}{n} \sum_k S_k Z_{n-k}.
\label{rec_proof} 
\eea

\nt
Where in the last two lines, we have simply redefined $n_k = n_k - 1$, which 
changes the sum to $\sum k n_k = n-k$. Since $k \leq 3$, for hadrons, 
we obtain the simplified recursion 
relation:

\bea
Z_n &=& \frac{1}{n} \Bigg[ S_1 Z_{n-1} + 2 S_2 Z_{n-2} + 3 S_3 Z_{n-3} \Bigg]. \label{recur1}
\eea

\begin{figure}[htb!]
\begin{center}
\hspace{0cm}
  \resizebox{4in}{3in}{\includegraphics[0in,1in][8in,9in]{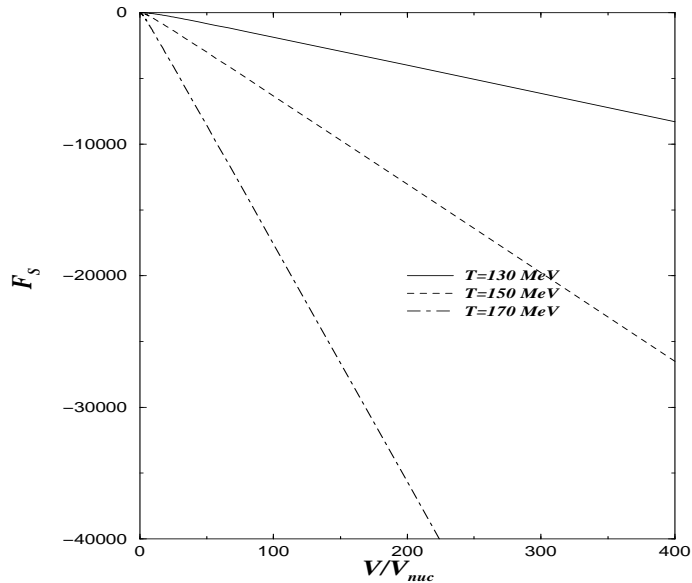}} 
    \caption{ Strange free energy $F_s =  -T \log \left( \frac{Z}{Z_0} \right)$
     as a function of temperature and thermalization volume. The 
     $x$-axis is the volume of the domain as a multiple of the 
     volume of a nucleon. }
    \label{free1}
\end{center}
\end{figure}

The actual method of computation involves simply calculating $S_1,S_2,S_3$,
and using this to calculate $Z_1, Z_2 , Z_3$ and so on in ascending 
order. A similar method may be used to compute the $Z_{-n}$'s. 
The presence of the $n!$ in the denominator of $Z_n$ 
(see Eq. (\ref{gnrl_part})) greatly reduces the magnitude of $Z_n$ for large $n$.
As a result, one never needs to go beyond $n=200$ to obtain an 
accuracy of 12 significant digits. The use of the recursion relations,
reduce such computations to a fraction of a second on a regular PC. Having 
calculated the partition function we may easily compute the 
strange free energy $F_s$, where

\bea
F = -T \log (Z) = -T \log \left( Z_0 Z_S \right) = F_0 + F_s.
\eea
  
\nt This is plotted in Fig. (\ref{free1}) as a function of 
domain size and temperature.

%%%%%%%%%%%%%%%%%%%%%%%%%%%%%%%%%%%%%%%%%%%%%%%%%%%%%%%%%%%
%%%%%%%%%%%%%%%%%%%%%%%%%%%%%%%%%%%%%%%%%%%%%%%%%%%%%%%%%%%

\subsection{Moments of the distribution.}

%%%%%%%%%%%%%%%%%%%%%%%%%%%%%%%%%%%%%%%%%%%%%%%%%%%%%%%%%%%
%%%%%%%%%%%%%%%%%%%%%%%%%%%%%%%%%%%%%%%%%%%%%%%%%%%%%%%%%%%

\begin{figure}[htb!]
\hspace{-1cm}
\resizebox{3.25in}{3in}{\includegraphics[0in,1in][8in,9in]{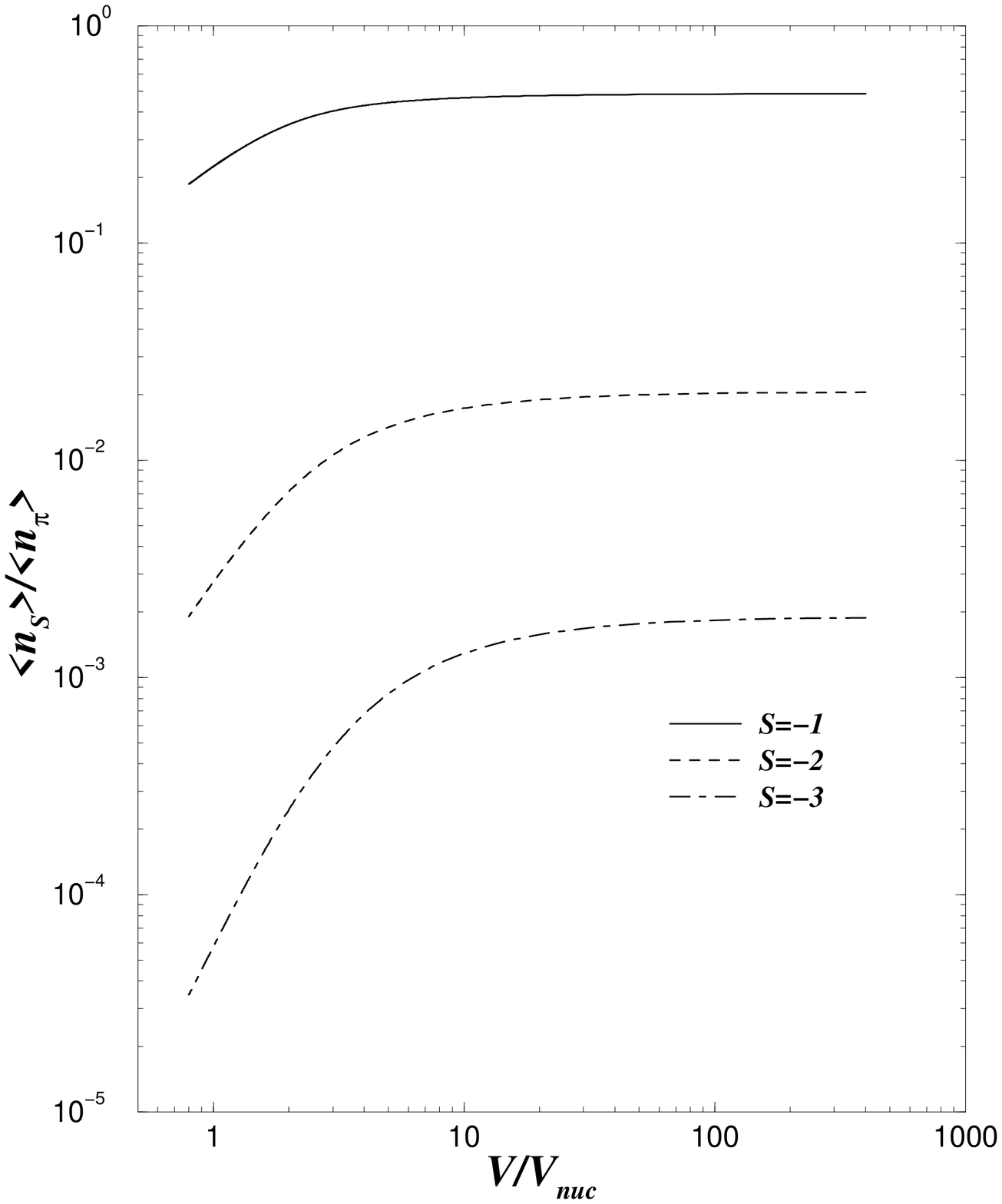}}  
 %\resizebox{16pc}{!}{\includegraphics[height=.2\textheight]{n_1_T130_mu0.ps}}
%\hspace{1cm}
\resizebox{3.25in}{3in}{\includegraphics[0in,1in][8in,9in]{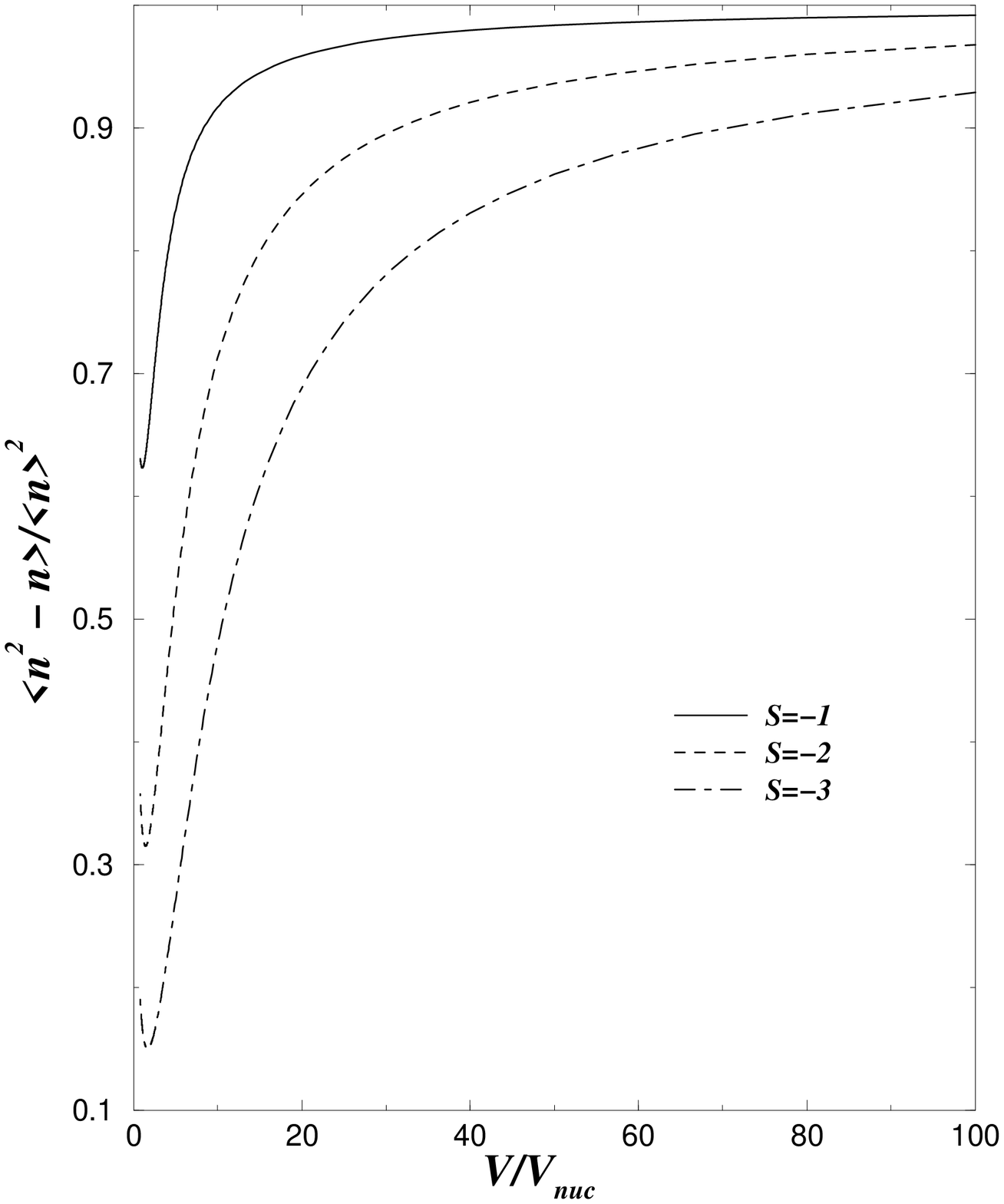}}
 %   \resizebox{16pc}{!}{\includegraphics[height=.2\textheight]{n2.ps}}
\caption{ The left panel plots the ratio of the mean number of strange 
hadrons to pions for hadrons with different strangeness. The right panel 
plots the quantity $f_2 = \lc n^2  - n \rc / \lc n \rc^2$. The legend is the 
same as for the left panel. Both plots are at a $T=170$MeV and $\mu=0$.}
    \label{der1}
\end{figure}

The recursion relations introduced in the previous subsection
may also be suitably modified to calculate the moments of the distribution. 
Formally these correspond to various derivatives 
of the partition function for different numbers of particles 
and may be calculated numerically as such. 
However, they may be exactly evaluated without the use of any numerical 
differentiation techniques.  In the following we simply illustrate the
calculation of the first and second moments. The higher moments
may be calculated in similar fashion. We will, only demonstrate
the method for strange hadrons; the calculation for 
anti-strange hadrons is almost identical. 

We focus on the general term in the series in Eq. (\ref{Z_series}). 
Expanding the strange part of this term, one obtains,

\bea
Z_0 Z_n Z_{-n} = Z_{0} \Bigg[ \sum_{n_1,n_2,...}   
\prod_k \frac{S_k^{n_k}}{n_k!} \Bigg] Z_{-n}.  
\eea

\nt
Each term in the sum represents the partition function of the 
system with the restriction that there be $n_1$ particles 
with strangeness 1, $n_2$ particles with strangeness 2 and so on. 
Dividing this by the full partition function will give us 
the probability to obtain this partition (we represent the set of 
numbers $n_1,n_2,...$ by the vector $\vn$),

\bea
P_{\vn} = \frac{ \left. \prod_k \frac{S_k^{n_k}}{n_k!} 
\right|_{\vn} Z_{-n}}{ \sum_n Z_n Z_{-n} }.    \label{prob}
\eea

\nt
The probability to obtain a particular partition, represents a 
central quantity in our computations. One may use this 
to compute any statistical quantity, \eg, 
the mean number of particles of strangeness $k$ will thus 
be given by the general expression,

\bea
\lc n_k \rc = n_k P_{\vn_1} + n_k P_{\vn_2} + ...
\eea

\nt
On the $r.h.s.$ of the above equation, $n_k$ is an operator acting on the 
partitions (or vectors) $\vn_1,\vn_2$ \etc\, 
The eigenvalue of this operator is the 
number of particles with strangeness $k$ 
in the vector $\vn_1,\vn_2$ \etc\, 
Substituting the value of $P_{\vn}$, we obtain,

\bea
\lc n_k \rc &=& \frac{ \sum_n \sum_{\vn} n_k \prod_i \left. \frac{S_i^{n_i}}{n_i!} 
\right|_{\sum_i i \, n_i = n} Z_{-n} }{ \sum_n Z_n Z_{-n} } \nn \\
&=& \frac{ \left. \sum_n \sum_{\vn}\right|_{\sum_{i=1}^{i_{max}} i \, n_i = n} 
\frac{S_1^{n_1}}{n_1!} ... \frac{S_k^{n_k}}{(n_k-1)!} 
... \frac{S_{i_{max}}^{ n_{i_{max}} } }{n_{i_{max}}!} Z_{-n} }{ \sum_n Z_n Z_{-n} } \nn \\
&=& \frac{ \sum_n \sum_{\vn} S_k \prod_i \left. \frac{S_i^{n_i}}{n_i!} 
\right|_{\sum_i i \, n_i = n - k} Z_{-n} }{ \sum_n Z_n Z_{-n} } .
\eea

\nt
Where, in the top line, 
the first sum represents the sum over different values of 
total strangeness, and the second sum represents the different 
partitions (or vectors) $\vn$ that will lead to the same total strangeness $n$. 
In the second line we have, for illustration, 
expanded the product $\prod_i$; here $i_{max}$ is equal to 3. 
In the last line we have, once again, shifted $n_k \ra n_k - 1$, and 
hence the sum $\sum_i i \, n_i$ is equal to $n - k$. Extracting $S_k$ from the 
sum, we note that the sum over all vectors $\vn$, under the
constraint $\sum_i i n_i = n - k$, is simply the pure strange partition
function with strangeness $n-k$, \ie,

\bea
\lc n_k \rc &=& \frac{\sum_{n=k}^{\infty}  S_k Z_{n-k} Z_{-n} }
{ \sum_{n=0}^{\infty} Z_n Z_{-n}} .\label{recur2}
\eea

\nt 
Using the same method as before to calculate the $Z_n$'s, we may simply 
calculate the mean number of particles with strangeness $k$. The 
ratio of the means for different strangeness to the number of pions is 
plotted in Fig. (\ref{der1}).

The second moment may also be easily evaluated. The natural 
quantity to evaluate in this recursion relation scenario is 
the term $f_2 \lc n \rc^2 = \lc n_k^2 - n_k \rc = \lc n_k(n_k - 1 ) \rc$, 
this is obtained from the general expression, 

\bea
\lc n_k^2 - n_k \rc = n_k (n_k - 1) P_{\vn_1} +  n_k (n_k - 1) P_{\vn_2} + ... 
\eea

\nt 
Substituting the expressions for the probabilities, and using methods 
almost identical to the derivation for the first moment, we obtain

\bea
\lc n_k(n_k - 1 ) \rc =  \frac{\sum_{n=2k}^{\infty}  S_k^2 Z_{n-2k} Z_{-n} }
{ \sum_{n=0}^{\infty} Z_n Z_{-n}} .\label{recur3}
\eea

\nt
The method of evaluation of these moments, involves simply 
evaluating the various $Z_n$'s to the required degree of accuracy and 
then using those to calculate the sums in the above expressions. 
We also plot the ratio of the second moment to the square of the 
first moment in Fig. (\ref{der1}). In all these plots, we 
have plotted the simplest case of zero chemical potential. 
It should be pointed out that 
incorporating a finite chemical potential involves almost no 
increase in the complexity or running time of the computation.

%%%%%%%%%%%%%%%%%%%%%%%%%%%%%%%%%%%%%%%%%%%%%%%%%
%%%%%%%%%%%%%%%%%%%%%%%%%%%%%%%%%%%%%%%%%%%%%%%%%
%%%%%%%%%%%%%%%%%%%%%%%%%%%%%%%%%%%%%%%%%%%%%%%%%

\section{The probability of $N$ particles from $p$ domains.}

%%%%%%%%%%%%%%%%%%%%%%%%%%%%%%%%%%%%%%%%%%%%%%%%%
%%%%%%%%%%%%%%%%%%%%%%%%%%%%%%%%%%%%%%%%%%%%%%%%%
%%%%%%%%%%%%%%%%%%%%%%%%%%%%%%%%%%%%%%%%%%%%%%%%%

In the previous sections we have dealt with the partition 
function, and strangeness production from a single domain.
In our picture, the central region in a heavy-ion collision 
has a full freeze-out volume $V_f$ which is divided into 
$p$ domains of volume $V$. In each domain, we have 
reached full thermal and chemical equilibrium in the canonical sense. 
As an aside, we point out here, as we did in the introduction that 
full chemical equilibrium may not have been reached, and the 
domain may in principle be chemically undersaturated or oversaturated. 
The usual 
method of dealing with such a scenario, is to introduce a 
fugacity factor $\gamma$ multiplicatively with the factors $S_k$. 
We leave this complication for a later effort. 
Yet another complication that will be ignored in this 
first effort, is a variation in the temperature and 
baryon chemical potential from domain to domain. 
All domains will be assumed to have the same temperature and 
chemical potential. In this section 
we will calculate the various probability distributions for 
the production of strange particles from the full system of volume $V$.
The reason for concentrating on the probability distribution is 
obvious: it represents the most basic quantity that may be 
computed in a given situation. All other quantities such as the 
mean, variance, higher moments 
\etc\, may be computed using the probability distributions.   

\begin{figure}[htb]
\hspace{-1cm}
  \resizebox{3in}{2.75in}{\includegraphics[0in,1in][6.5in,10in]{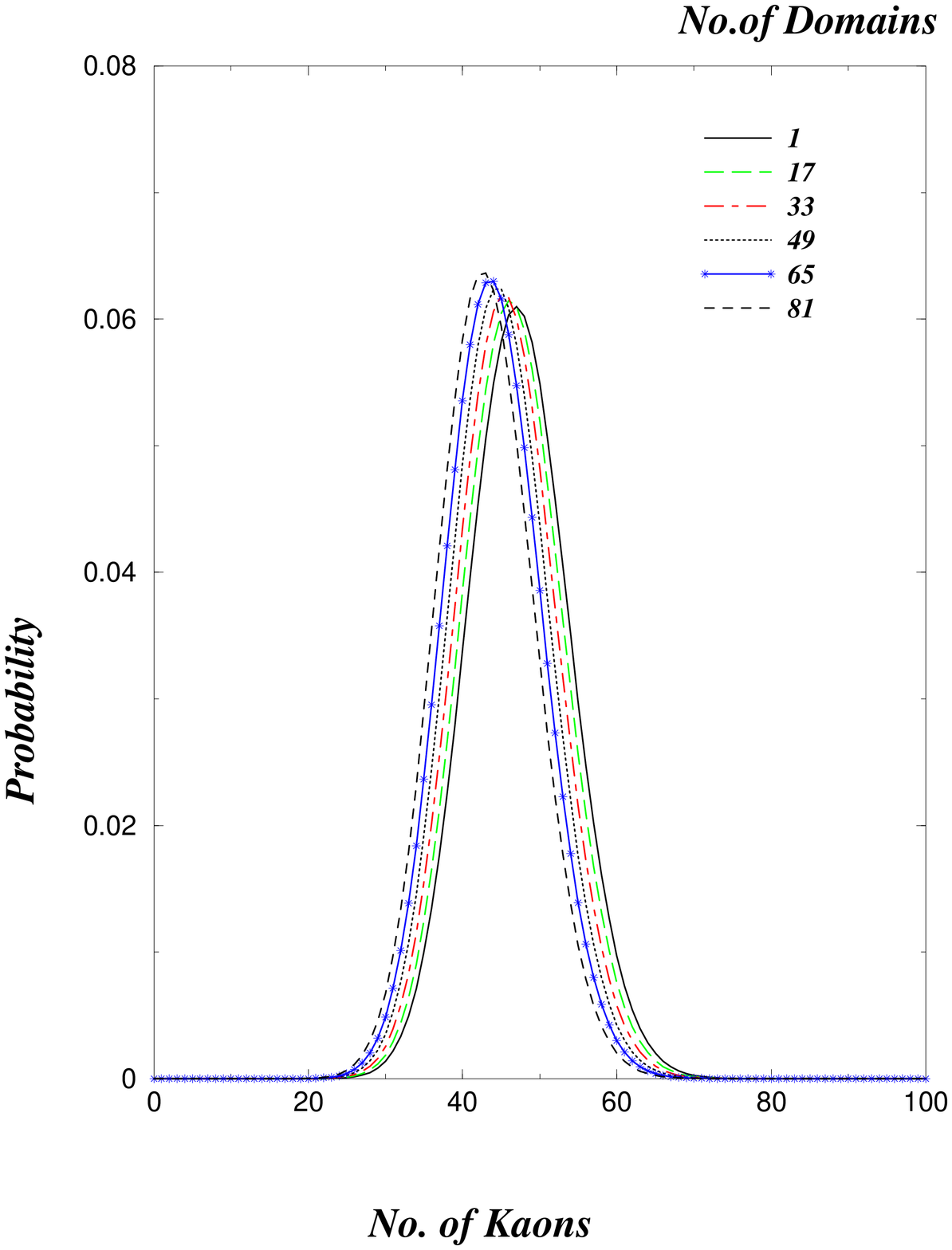}} 
\hspace{1cm}
  \resizebox{3in}{2.75in}{\includegraphics[0in,1in][6.5in,10in]{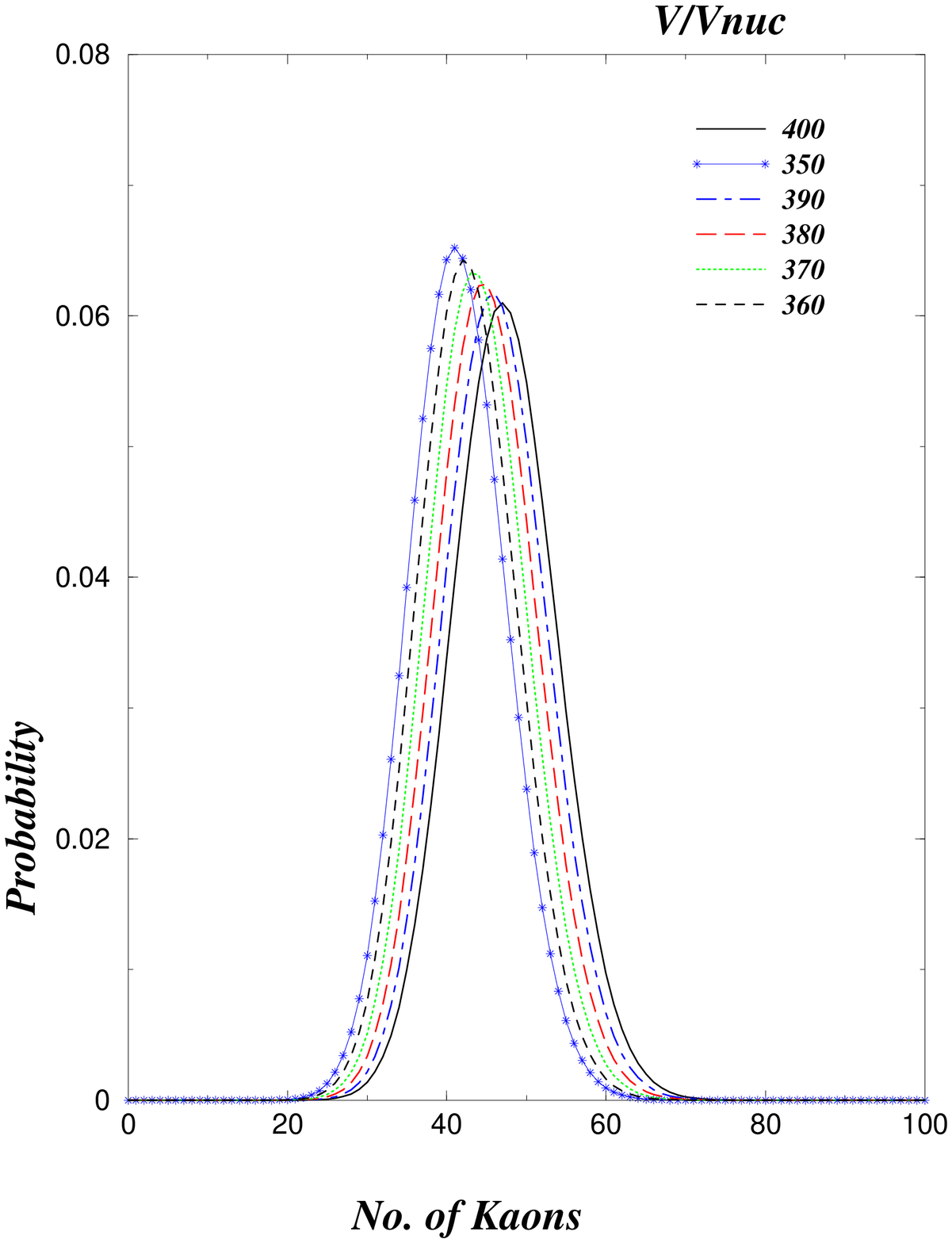}} 
    \caption{ Probability distributions for the production of 0 to 100 Kaons from a
    maximum total volume $V_f = 400 \, V_{nuc}$ ($V_{nuc}$ is the volume of a nucleon). 
    In the left panel the different distributions 
    represent the different scenarios of the 
    full volume divided into 1 to 81 independent domains. In the 
    right panel the different distributions indicate 
    different cases of total volumes from $400 \, V_{nuc}$ to $350 \, V_{nuc}$. }
    \label{probfig}  
\end{figure}

Of the variety of strange particles produced in a heavy-ion collision
the Kaon is by far the most common and the easiest to identify. We 
thus compute the probability to obtain $A$ Kaons from the full 
system that consists of $p$ domains, \ie, $P_p^A$. The 
method employed will be almost identical for particles with 
strangeness -2 ($\Xi$ particles) and -3 ($\Omega$ particles).  
The $A$ Kaons are indistinguishable bosons: thus the 
probability to obtain $A$ indistinguishable bosons from 
$p$ domains may be written schematically as

\bea
P_p^A = \frac{D_p^A}{A!},
\eea 

\nt
where $D_p^A$ is the probability to obtain $A$ distinguishable
Kaons from $p$ partitions. If there are $n_1$ Kaons emanating 
from domain 1, $n_2$ from domain 2, \etc, then we define the quantity,

\bea
D_p^A(n_1,n_2...n_p) = 
\frac{A!}{n_1! n_2! ... n_p!}  D_1^{n_1} D_1^{n_2} ... D_1^{n_3}.
\eea

\nt 
Using this we may express the probability to obtain $A$ distinguishable
Kaons from $p$ domains as,

\[
D_p^A = \sum_{n_1,n_2...n_p | \sum_i n_i = A} D_p^A(n_1,n_2...n_p).
\]

\nt As a result we obtain,

\bea
P_p^A &=&  \sum_{n_1,n_2,...n_p | \sum_{k=1}^p n_k = A} 
\frac{D_1^{n_1}}{n_1!} \frac{D_1^{n_2}}{n_2!} ... \frac{D_1^{n_p}}{n_p!}
\nn \\ 
&=& \sum_{n_1,n_2,...n_p | \sum_{k=1}^p n_k = A} 
P_1^{n_1} P_1^{n_2} ... P_1^{n_p} .  \label{fullprob}
\eea  

\nt
Where $P_1^{n_k} = \frac{D_1^{n_k}}{n_k!}$ 
represents the probability to obtain $n_k$ 
indistinguishable Kaons
from 1 domain. 
In Eq. (\ref{gnrl_part}) the $S_k$ represent the 
single particle partition functions for particles with strangeness 
$k$. There are many such particles. 
The quantity $S_k$ is composed of a sum of the single particle partition 
functions for each of these particles (see Eq. (\ref{sk})).     
From Eq. (\ref{sk}) we extract $S(K)$, and define the quantity $\bar{S}_1$, as

\[
S_1 = S(K) + S(K^*) + S(K_1) + S(\Lambda)  + S(\Sigma) + ... 
= S(K) + \bar{S}_1 .
\]

\begin{figure}[htb]
\hspace{-1cm}
  \resizebox{3in}{2.75in}{\includegraphics[0in,1in][6.5in,10in]{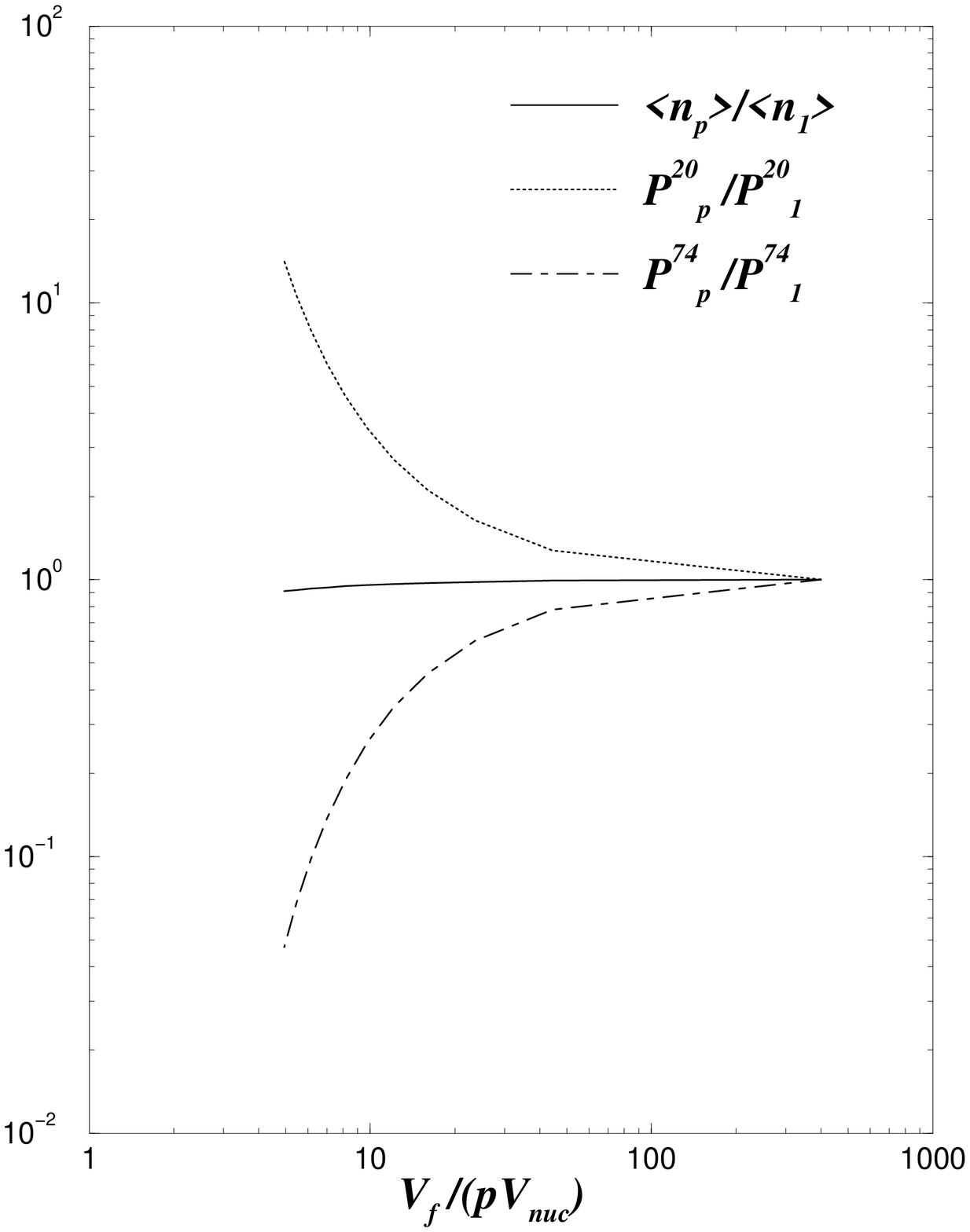}} 
\hspace{1cm}
  \resizebox{3in}{2.75in}{\includegraphics[0in,1in][6.5in,10in]{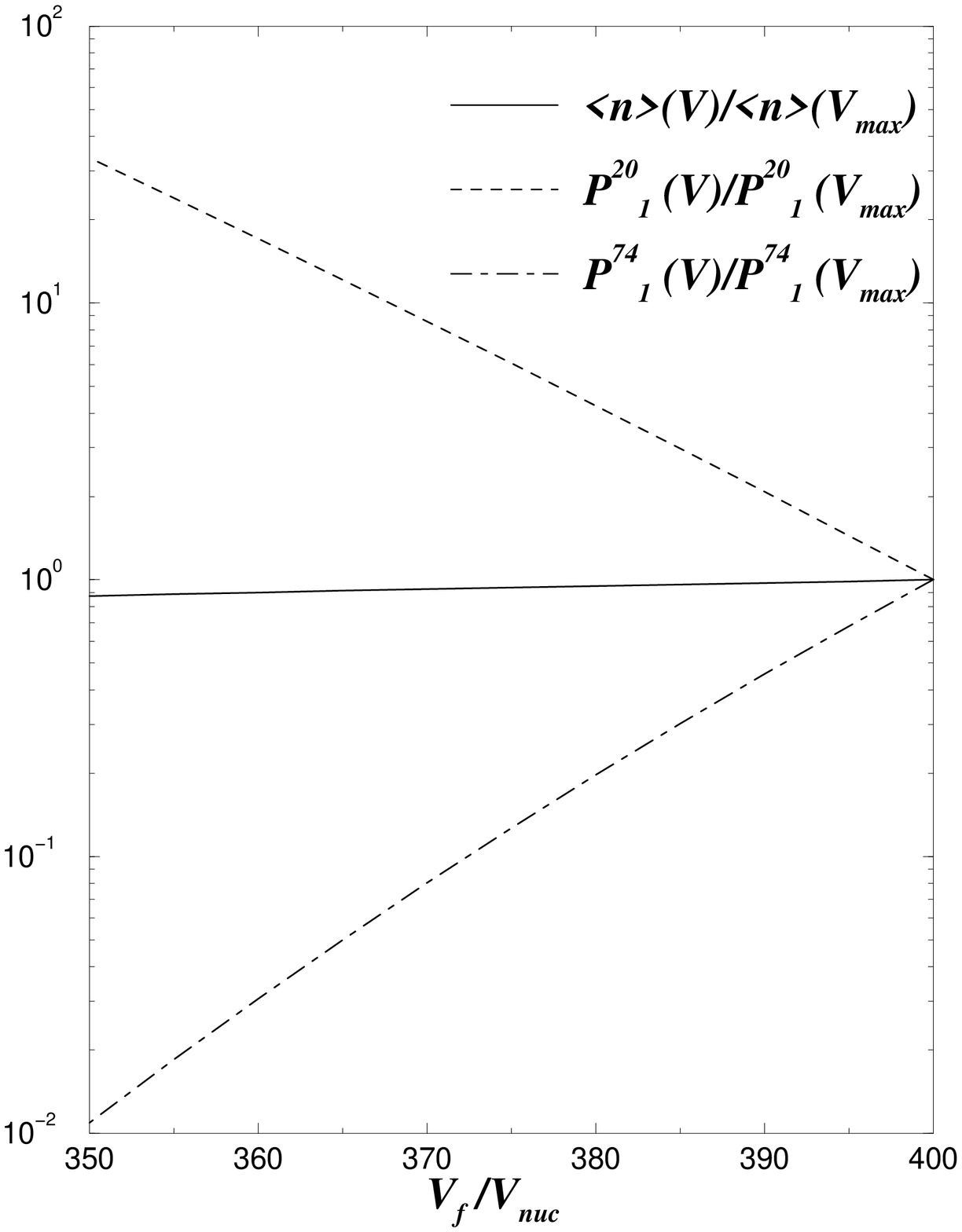}} 
    \caption{ Left panel shows the variation of the 
    mean number of Kaons, probability to 
    produce 20 and 74 Kaons from a volume of size $400 \, V_{nuc}$ as 
    the number of domains is increased. The variation with 
    the size of a domain is plotted. All three quantities are normalized 
    to their values for an unpartitioned volume. The right panel 
    plots the variation of the same three quantities from an 
    unpartitioned decreasing volume. All quantities are normalized to 
    their values at $V=400 \, V_{nuc}$.  }
    \label{avefig}  
\end{figure}

With this, we define a partition function with total strangeness $n$,
consisting of only strange particles but not containing any Kaons, as

\bea
\bar{Z}_n = \sum_{n_1,n_2,... | \sum k n_k = n}  \frac{ \bar{S}_1^{n_1}}{n_1!}
\frac{ S_2^{n_2}}{n_2!} \frac{ S_3^{n_3}}{n_3!}.
\eea     

\nt
The above partition function, may also be easily evaluated with a similar 
recursion relation:

\[
\bar{Z}_n = \frac{1}{n} \Bigg[ \bar{S}_1 \bar{Z}_{n-1} + 
2 S_2 \bar{Z}_{n-2} + 3 S_3 \bar{Z}_{n-3} \Bigg].
\]

\nt 
In the above $S_2$ and $S_3$ remain unaffected. Using the above 
two equations one may construct the probability $P_1^n$ of 
obtaining exactly $n$ Kaons from a single domain, with net 
strangeness zero. Using Eq. (\ref{prob}), this may be written as,

\bea
P_1^n = \frac{Z_0}{Z} \frac{S(K)^n}{n!} 
\sum_{N=0}^{\infty} \Bigg[  \sum_{n_1,n_2,n_3 | \sum i n_i = N} 
\frac{ \bar{S}_1^{n_1} }{n_1!} \frac{ S_2^{n_2} }{n_2!} \frac{ S_3^{n_3} }{n_3!} 
\Bigg] Z_{-(N+n)} .
\eea

\begin{figure}[htb]
\hspace{-1cm}
  \resizebox{3in}{2.75in}{\includegraphics[0in,1in][6.5in,10in]{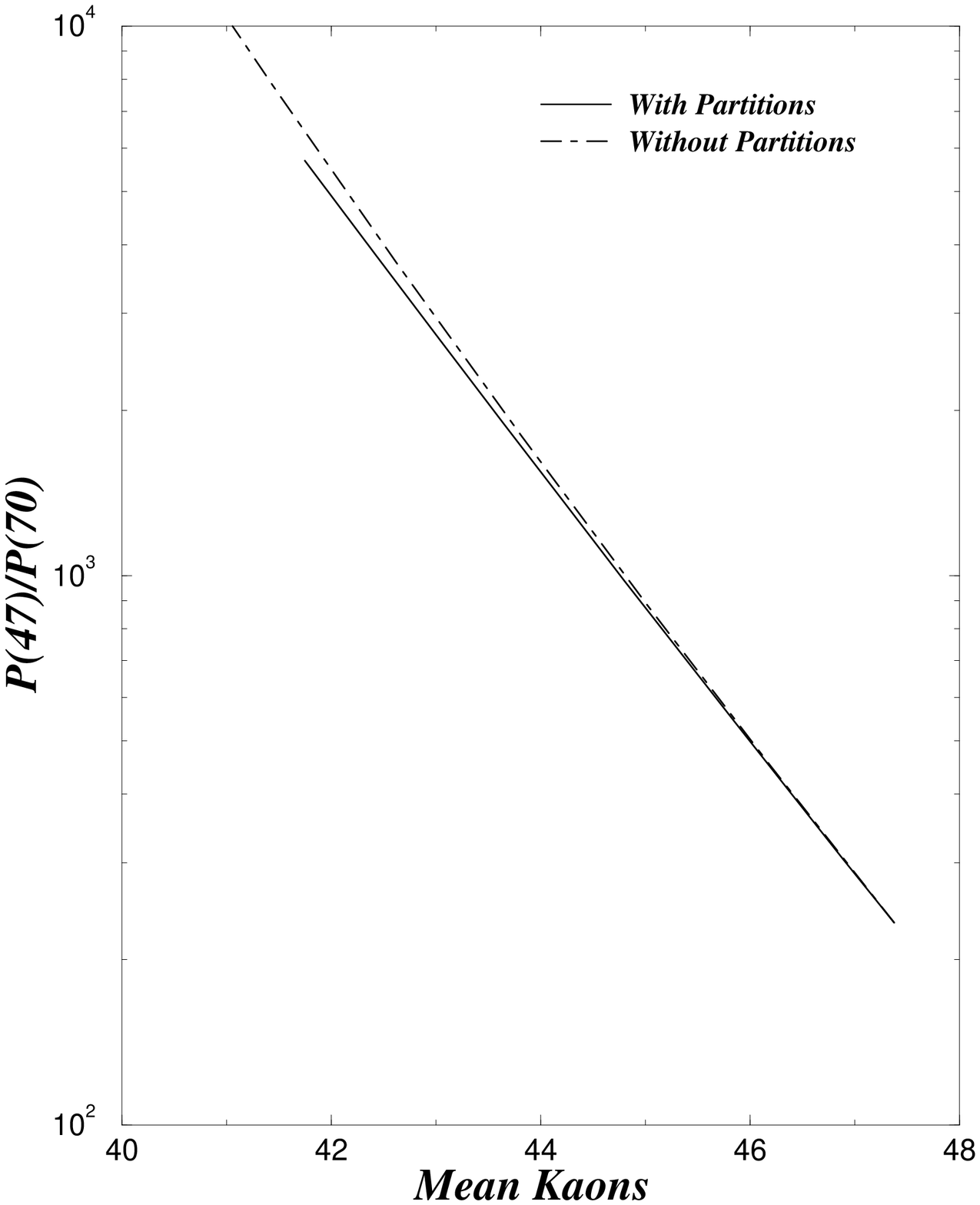}} 
\hspace{1cm}
  \resizebox{3in}{2.75in}{\includegraphics[0in,1in][6.5in,10in]{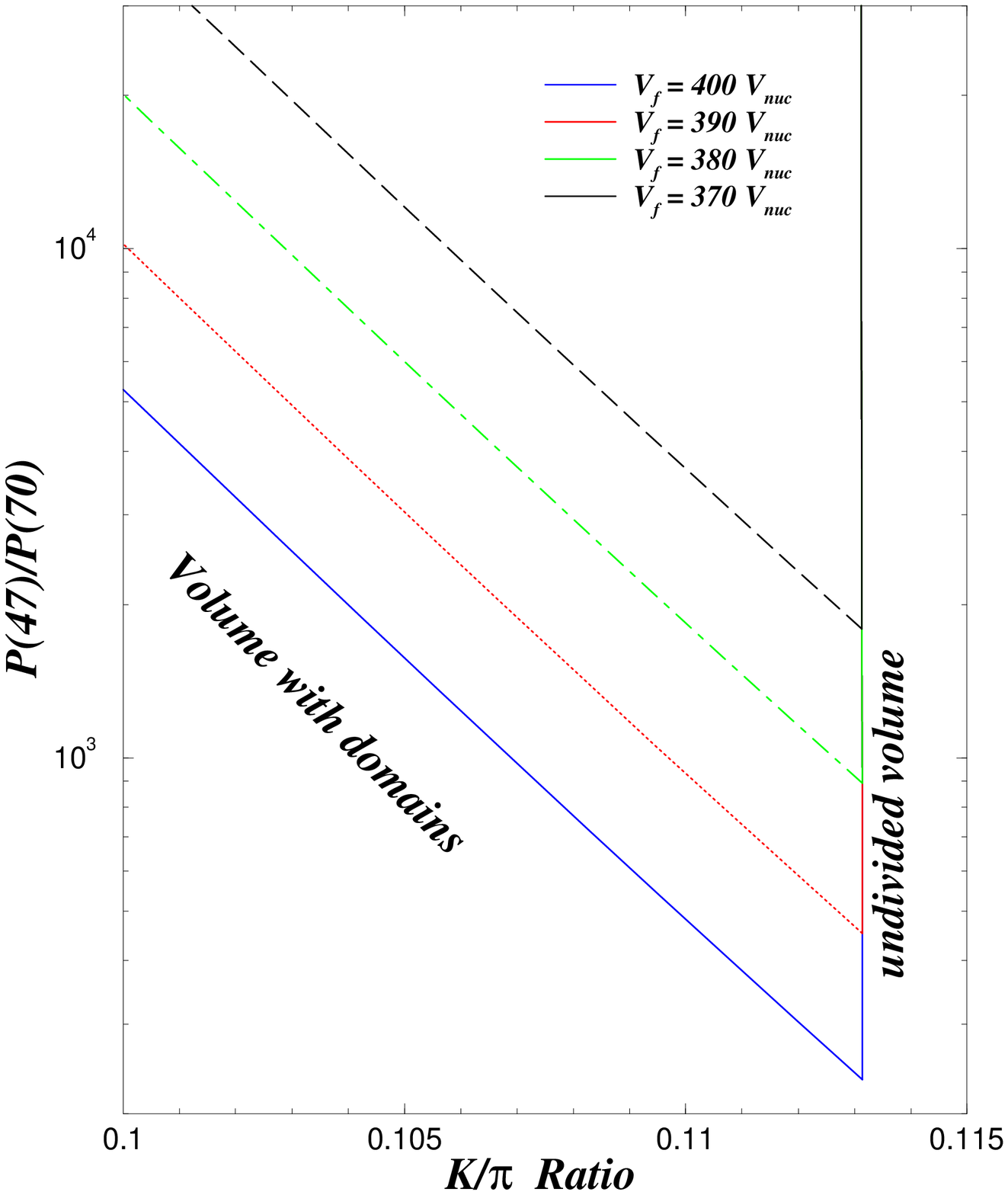}} 
\caption{ Left panel shows the variation of the the ratio of probabilities 
    $(P(47)/P(70))$ as a function of the mean number of Kaons. 
    The two quantities change as the number of domains is increased, or the 
    total volume is decreased. The right panel shows the variation of the 
    same ratio of probabilities as a function of the $K/\pi$ ratio. The 
    vertical line represents the case of decreasing the total 
    unpartitioned volume. The slanted lines represent the case of 
    increasing the number of domains in a given total volume. }
\label{ratfig}  
\end{figure}

In the above equation, the term in square brackets represents the partition 
function for a box containing only strange particles, with net 
strangeness $N$, none of which are Kaons. Obviously, the term in 
square brackets is simply $\bar{Z}_N$. With this we may write down the full 
probability to obtain $A$ Kaons from the full system with $p$ partitions as

\bea
P_p^A &=& \sum_{n_1,n_2,...n_p | \sum_k n_k = A} \left( \frac{Z_0}{Z} \right)^p 
\prod_{k=1}^p \left( \frac{ S(K)^{n_k} }{n_k!} \sum_{t_k = 0}^{\infty} 
\bar{Z}_{t_k} Z_{-(t_k + n_k)}  \right)  \nn \\
&=& \left( \frac{Z_0}{Z} \right)^p S(K)^A \sum_{n_1,n_2,...n_p | \sum n_k = A} 
\prod_k w_{n_k}. \label{probsum}
\eea 

\nt 
In the above equation 

\bea
w_{n_k} = \left( \sum_{t_k = 0}^{\infty} 
\bar{Z}_{t_k} Z_{-(t_k + n_k)} \right) \Bigg/ n_k! 
\eea

\nt
Where $w_{n_k}$'s for $0 < n_k < A$ 
may be easily calculated using the $\bar{Z}_t$ and the $Z_{-t}$'s. The 
evaluation of the sum with the constraint in Eq. (\ref{probsum}) seems 
to be extremely complicated. However, yet another recursion relation 
exists which allows this to be computed swiftly. Note that the sum 

\[
S_p^A = \sum_{n_1,n_2,...n_p | \sum n_k = A} 
\prod_k w_{n_k} ,
\] 

\nt
requires us to distribute $A$ particles in $p$ distinguishable boxes. 
If we put $q$ particles in the first box, then we have to 
distribute $A-q$ particles in $p-1$ boxes. Summing over $q$ gives us 
the recursion relation, 

\bea
S_p^A = \sum_{q=0}^A w_q S_{p-1}^{A-q}.  \label{lastrecur}
\eea
 
\nt 
One may make the simple observation that $S_1^k = w_k$. This along with 
Eq. (\ref{lastrecur}) allows to calculate the probability $P_p^A$ recursively.
One essentially starts with $S_1^k$, uses these to build the $S_2^k$ and 
so on, until one obtains $S_p^A$ with the chosen values of $p$ and $A$.    
A plot of these probabilities is presented in
Figs. (\ref{probfig},\ref{probc},\ref{probw}).

%%%%%%%%%%%%%%%%%%%%%%%%%%%%%%%%%%%%%%%%%%%%%%%%%
%%%%%%%%%%%%%%%%%%%%%%%%%%%%%%%%%%%%%%%%%%%%%%%%%
%%%%%%%%%%%%%%%%%%%%%%%%%%%%%%%%%%%%%%%%%%%%%%%%%

\section{Results and discussions}

%%%%%%%%%%%%%%%%%%%%%%%%%%%%%%%%%%%%%%%%%%%%%%%%%
%%%%%%%%%%%%%%%%%%%%%%%%%%%%%%%%%%%%%%%%%%%%%%%%%
%%%%%%%%%%%%%%%%%%%%%%%%%%%%%%%%%%%%%%%%%%%%%%%%%

\begin{figure}[htb]
\hspace{-1cm}
  \resizebox{3in}{2.75in}{\includegraphics[0in,1in][6.5in,10in]{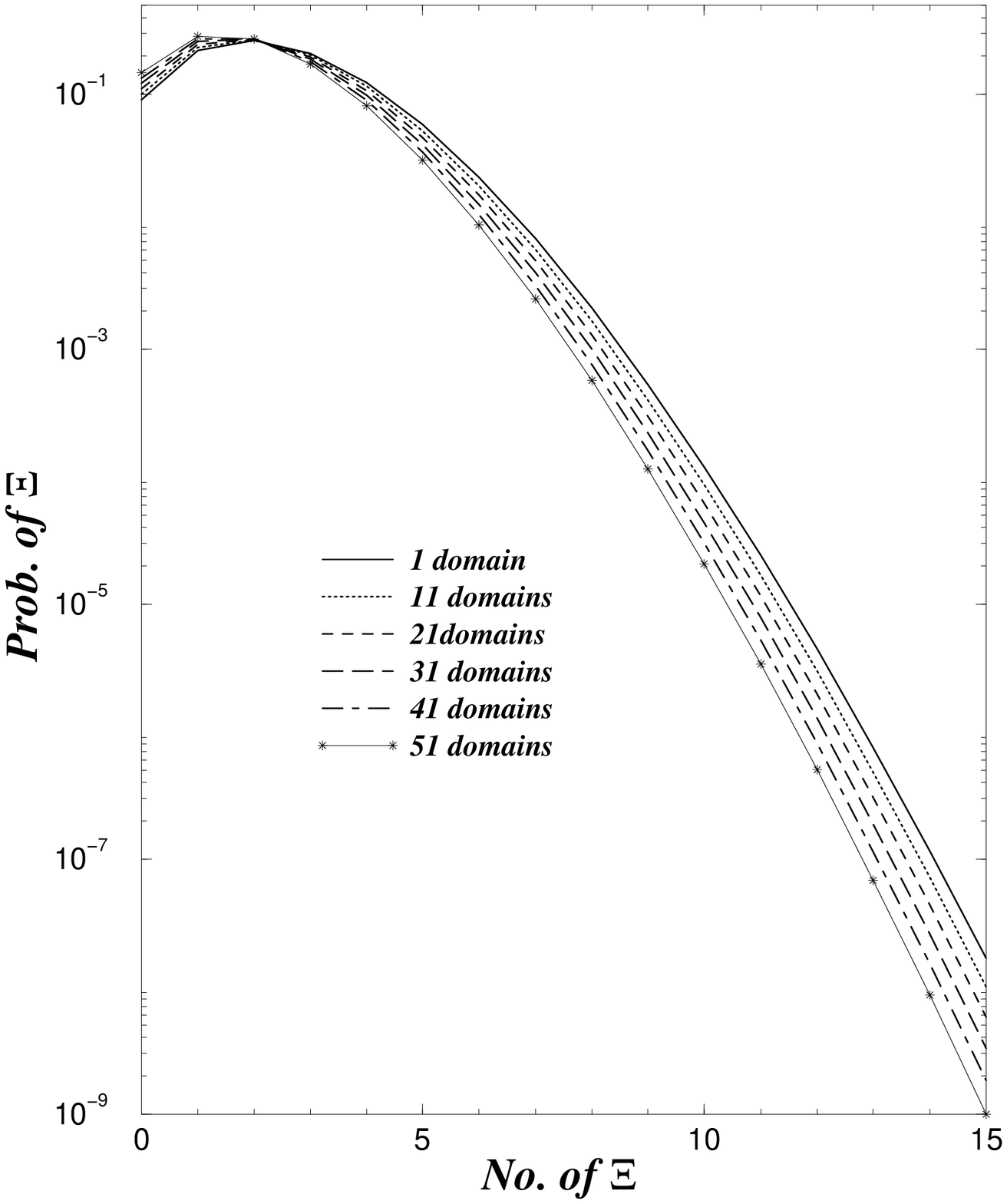}} 
\hspace{1cm}
  \resizebox{3in}{2.75in}{\includegraphics[0in,1in][6.5in,10in]{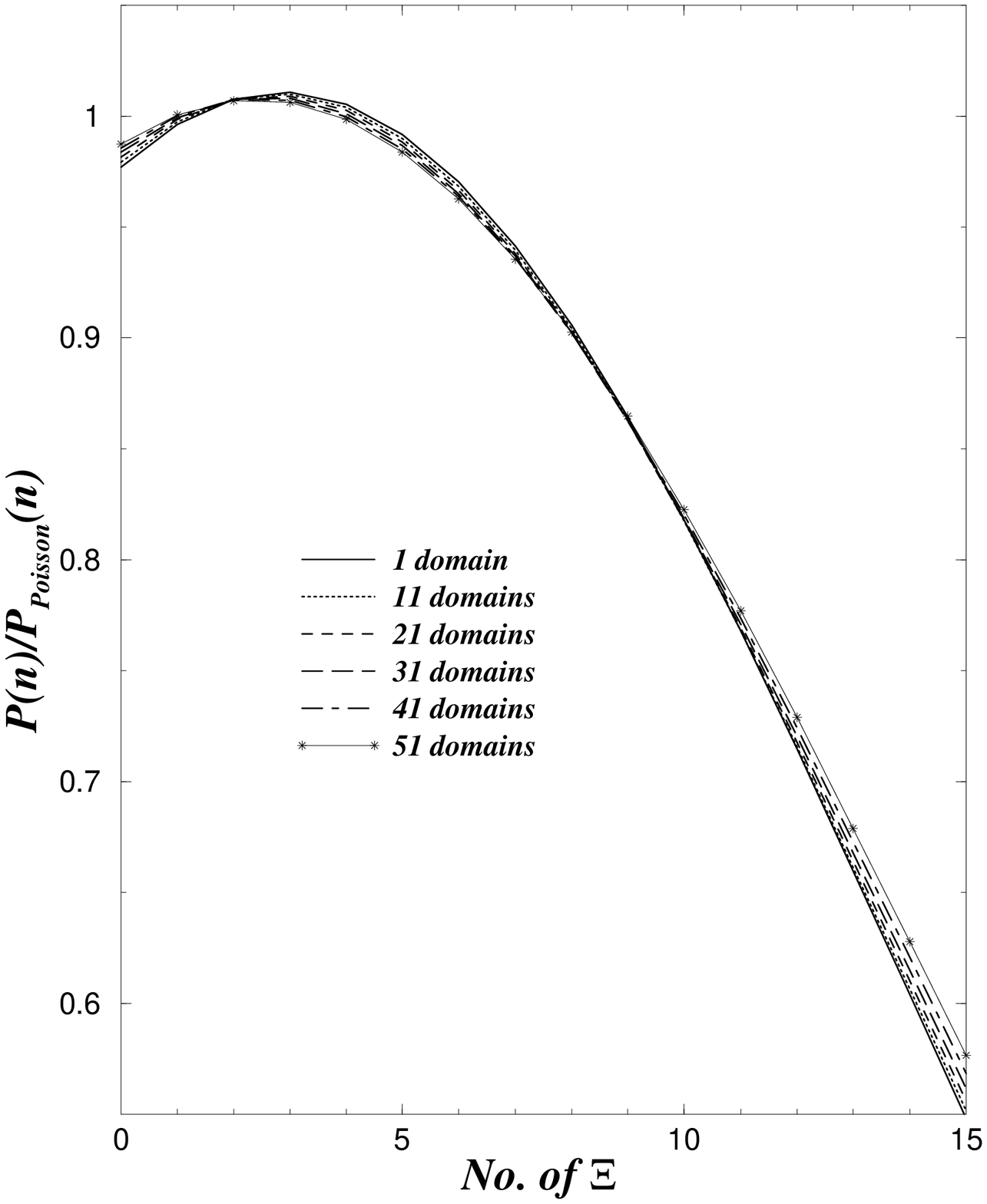}} 
\caption{ The left panel shows the probability distributions of  
    $\Xi$ particles for a full $V_f = 400 \, V_{nuc}$ and $T=170$ MeV. The volume is 
    divided into various numbers of partitions 1,11,21,31,41,and 51 as 
    shown. The distributions shift slowly towards the left. The mean 
    for each distribution is calculated and is used to compute a 
    Poisson distribution (details in text). The ratio of the 
    calculated probability distributions to the corresponding 
    Poisson distributions is then plotted in the right panel.}
    \label{probc}  
\end{figure} 

We begin the analysis with the kaon distributions (Fig. (\ref{probfig})). 
In our calculations, the full freeze-out volume $V_f$ 
is taken to be about 400 times the volume
of a nucleon. In the left panel of Fig. (\ref{probfig}), 
various cases are presented where this $V_f$ is 
divided into 1,17,33...81 domains. In comparison, 
the right panel represents the probability distributions for one 
partition, but for different cases of a decreasing full freeze-out volume. 
The volumes range from $V_f = 400 \, V_{nuc} $ to $V_f = 350 \, V_{nuc} $.
Note that, in the left panel, as the number of domains is 
increased, the Gaussian like curves slowly shift toward the left. 
A similar pattern is realised if instead of 
partitioning the volume, we simply reduced the full volume by 10\%, 
constraining the system to only one domain.
Although the probability distributions vary with the partitioning of the system 
as well as the 
reduction of the volume, a concise statement about the change, or even the 
difference between the two cases may not 
be made from Fig. (\ref{probfig}). 
A clearer picture of the problem  
may be obtained from Fig. (\ref{avefig}): 
here we plot the mean number of kaons $\lc n_k \rc$ 
as the number of domains is increased (left panel), and as the the full freeze-out volume 
is decreased (right panel). For an unpartitioned volume with $V_f = 400 \, V_{nuc}$ the 
mean $\lc n \rc \simeq 47$ Kaons. 
Along with this, we also plot the probability to get 
$\lc n_k \rc + 27$ and $\lc n_k \rc - 27$ kaons. These 
probabilities sample the edges of the probability 
distributions, which we believe will be most sensitive 
to a change of the macroscopic conditions. 
In the following, we will denote these as ``edge'' probabilities. 
For an unpartitioned volume with $V_f = 400 \, V_{nuc}$ these 
``edge'' probabilities represent the probabilites to obtain 74 and 20 
Kaons. The values of such probabilities are $\sim 10^{-6}$ and are 
measurable in experiments. One may go even further towards the edges of 
the distribution to observe a larger effect of partitioning the system, 
however the values of such probabilities may be too small to be measurable in
experiments. 
In the figure, all
values are normalized to the values of the quantities obtained
from the unpartitioned maximum volume ($V_f=400 V_{nuc}$).
Note that the mean hardly changes as we increase 
the number of domains, or reduce the volume. The ``edge'' 
probabilities however display substantial variation.
In a real heavy-ion collision, one can never determine the 
volume to a great degree of accuracy
(in the case of flow, the very meaning of
the volume becomes complicated). Hence, the similarity of the 
two plots makes it difficult to distinguish between the two cases. 
If only the kaon probability distributions were measured, sole observation 
of these would not allow one to distinguish between these two cases. 
Measurements of other observables, not constrained by 
strangeness conservation (such as pion yields), 
will have to be invoked.

Even though the ``edge'' probabilities in the left and 
right panels of Fig. (\ref{avefig}) look substantially different, 
one cannot simply superpose the two figures to make a comparison, 
the $x$-axis in 
both graphs represent very different quantities.
One may use a limiting procedure by calculating the absolute 
number of particles not constrained by conservation laws, 
such as pions. These are sensitive to 
the total volume and can be used to set constraints 
on the total volume. 
This limits the range of the $x-$axis of the right panel of Fig. (\ref{avefig}).
Such a constraint will set limits on the maximum spread between the two 
``edge'' probabilities. 
If the spread between the 
``edge'' probabilities in the left panel 
went beyond the limits set by the right panel, then one would be 
lead to the conclusion that a large thermalized 
system was not formed. Thus, to determine the amount of 
partitioning of the total volume, one would need to determine 
the pion or proton yields with an error somewhat smaller than the 
range of the $x$-axis of Fig.(\ref{avefig}). Thus, discering the 
degree of thermalization will require accurate experimental results.

\begin{figure}[htb]
\hspace{-1cm}
  \resizebox{3in}{2.75in}{\includegraphics[0in,1in][6.5in,10in]{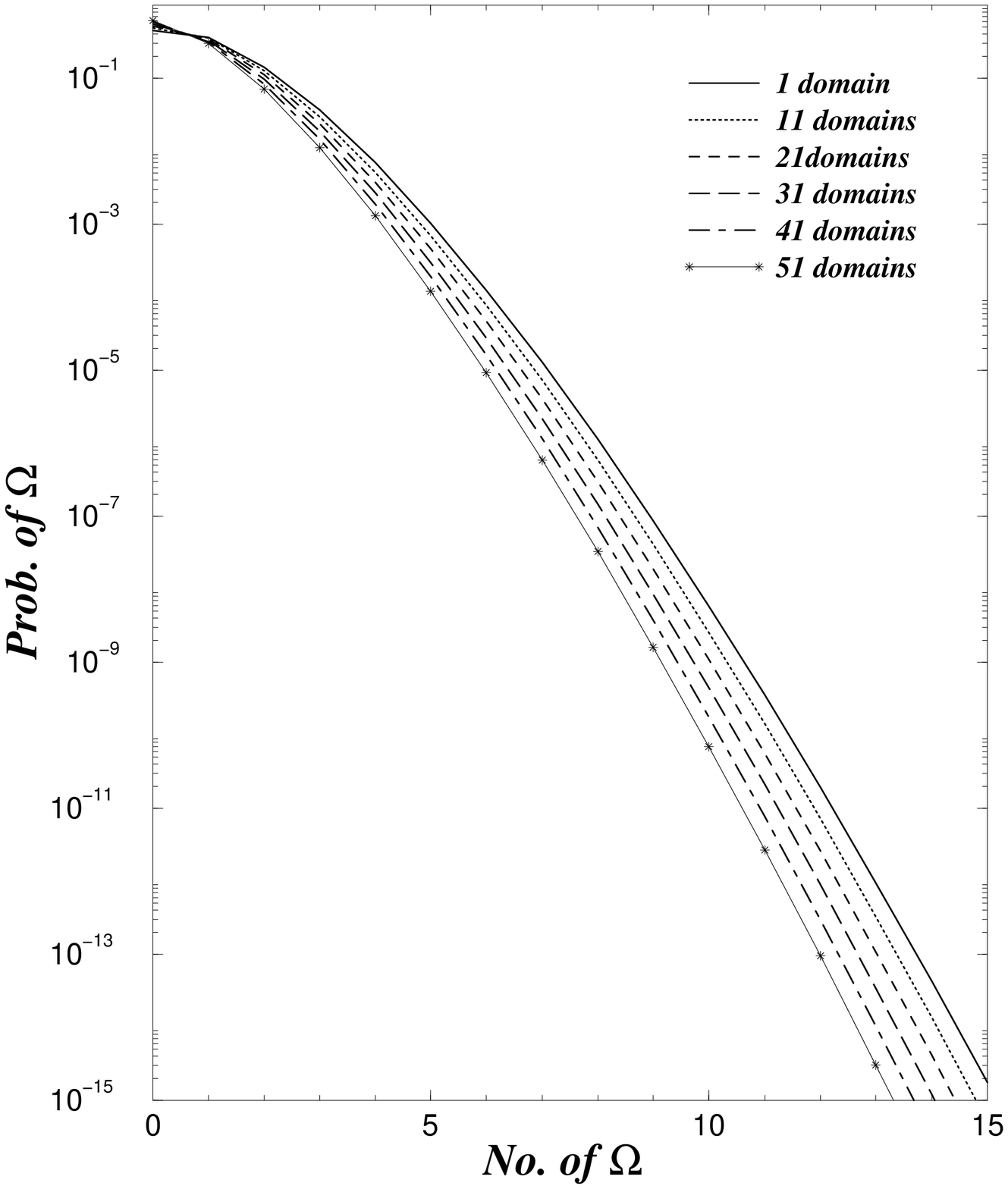}} 
\hspace{1cm}
  \resizebox{3in}{2.75in}{\includegraphics[0in,1in][6.5in,10in]{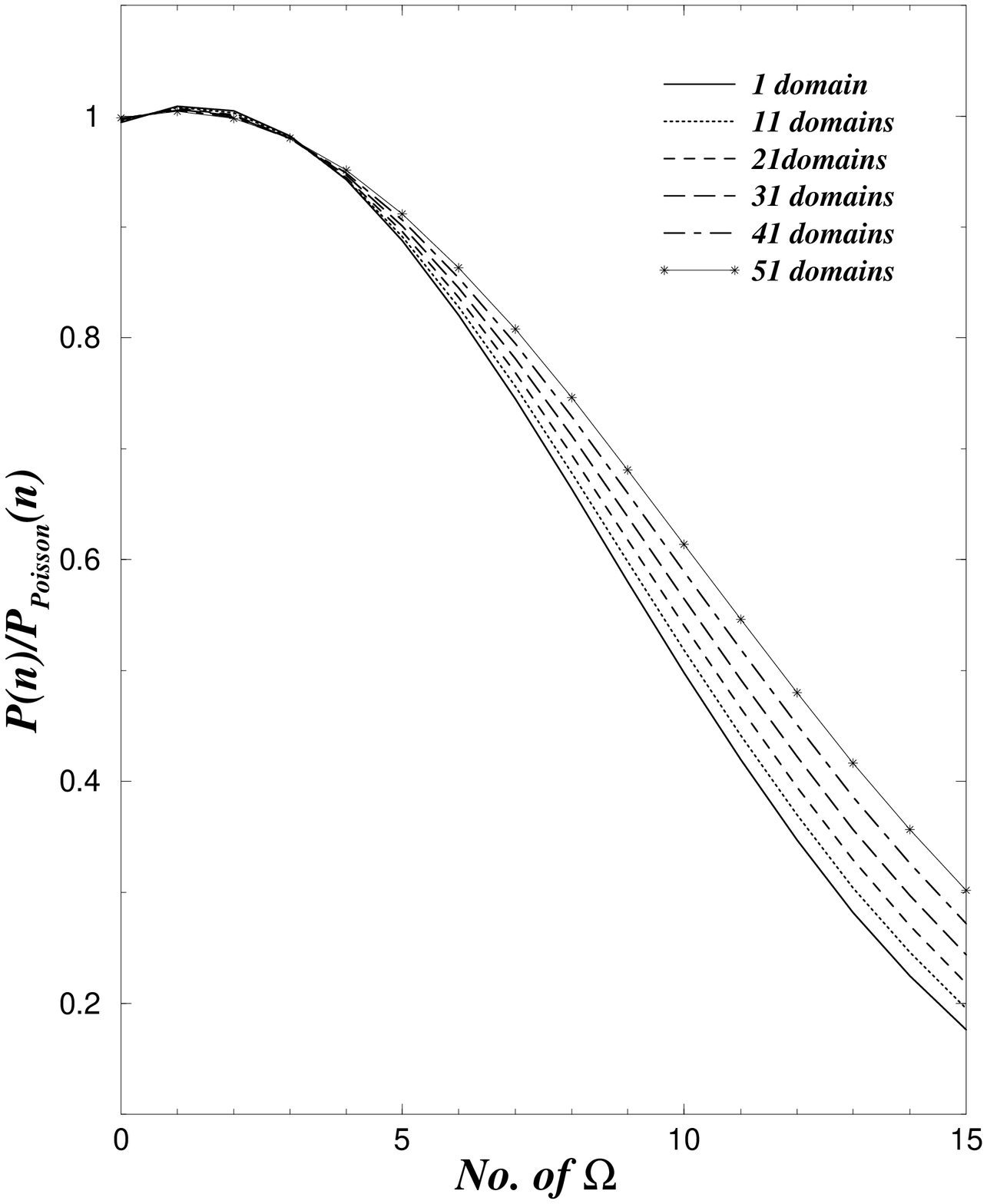}} 
    \caption{ Same as Fig. (\ref{probc}) but for the $\Omega$ particles.}
    \label{probw}  
\end{figure}

As mentioned before, a change in the total volume would 
lead to a noticeable change in the pion number. The 
kaon-to-pion ratio, however, would remain almost constant in the 
case of a single domain, as both are proportional to the volume.
The behavior of this ratio versus 
the volume, for a single domain, is plotted in Fig. (\ref{der1}). 
In Fig. (\ref{ratfig}) (right panel), we plot a ratio of probabilities 
versus the $K/\pi$ ratio from both scenarios. 
The choice of which two probabilities to use is arbitrary.
We chose the numerator near the peak of the distribution $(P(47))$,
while the denominator is chosen near the edge $(P(70))$. 
In the following, this will be denoted as the ``peak-edge'' ratio. 
The motivation behind this choice is that as the probability 
distributions (in Fig. (\ref{probfig})) shift towards the left, 
the value near the peak 
changes marginally in comparison to the value at the edge, 
which drops rapidly. 
%Here we plot the ``peak-edge'' ratio for kaons 
%versus the $K/\pi$ ratio. 
Experimentally, this will 
correspond to the case where the temperature $T$ has been measured from 
other ratios. This temperature $T$ is then used to plot the curves in 
Fig. (\ref{ratfig}). 
The vertical line represents the case 
of particle production from an unpartitioned volume. The $K/\pi$ 
ratio is almost a constant as expected. The ``peak-edge'' ratio  
simply grows as the volume is reduced. 
The slanted set of lines represents 
the case for multiple domains. The $K/\pi$ ratio drops as the 
number of domains increases. As the total volume is unchanged, 
the number of pions remains fixed. The different lines represent 
different cases where we have considered smaller freeze-out volumes $V_f$. 
The cases plotted are for $V_f = 400, \, 390, \, 380, \, 370 \, V_{nuc}$. 
Thus, the triangular region represents the ``allowed region'' where an 
experimental point may lie if such an observation were made. 
The location of the vertical line of the region depends 
solely on the temperature. 
The location of the actual experimental point allows one to 
decipher not only the total volume which contributes to the production 
of particles but also the size of the domain over which 
strangeness is conserved. 
%This provides a lower limit on 
%the size of the region over which kinetic equilibrium is achieved.
However, as before, we note that, even though the ``peak-edge'' ratio 
varies over many orders of magnitude; the $K/\pi$ ratio shows only a 
15 \% variation. Thus, both the $K/\pi$ ratio and $T$ have 
to be determined with an error 
much less than 15 \%. 
This reinstates our earlier result that  
accurate experimental results are required to quantitatively 
describe the degree of thermalization.

The importance of an accurate measurement of the pion yield may be
further underlined if we plot the 
``peak-edge'' ratio versus only the kaon yield, this is 
done in the left panel of Fig. (\ref{ratfig}).  
The solid line represents the variation of the ``peak-edge'' 
ratio of probabilities
with the mean number of Kaons emanating from multiple
domains in a full volume of $400 \, V_{nuc}$. The dot-dashed line 
represents the same comparison but for a total volume 
descending from a maximum value of $400 \, V_{nuc}$. Note that
both curves lie almost on top of one another.  
The ``peak-edge'' ratio drops slightly more sharply for the partitioned 
case. The main reason for the two plots being similar is, 
not an ill choice of the probability ratio, but rather, the mean number 
of kaons, which varies almost linearly with change in volume for 
volumes in the range of $350-400 \, V_{nuc}$. This allows the kaon 
number to change by almost the same range in both plots. If 
we were to proceed to a much smaller number of kaons the two graphs 
would then separate from each other. This demonstrates, yet again, 
the similarity of the probability distributions in Fig. (\ref{probfig}). 
%and also the need for an accurate determination of the pion yield.

\begin{figure}[htb]
\begin{center}
\hspace{0cm}
  \resizebox{4in}{3.75in}{\includegraphics[0in,1in][8in,10in]{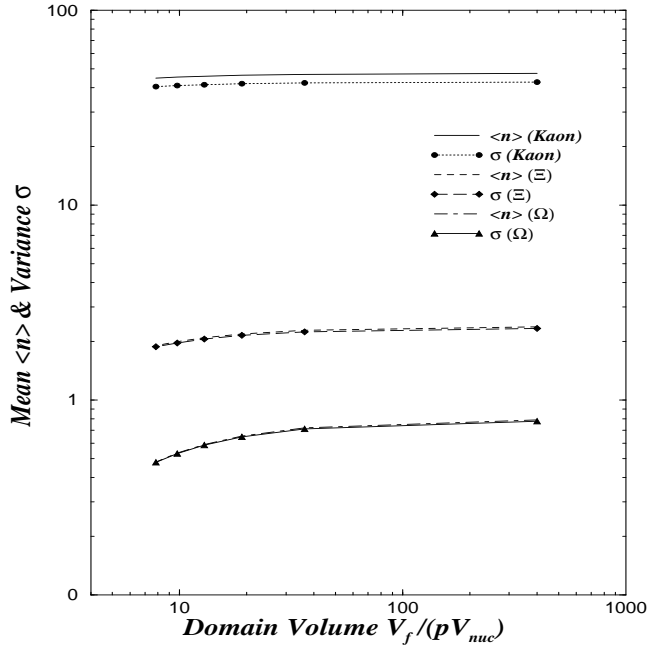}} 
    \caption{ Mean and Variance for 
    Kaons, $\Xi$ and $\Omega$ particles for a $T=170$MeV,
    and a total $V=400 \, V_{nucleon}$. 
    The number of domains varies from 1 to 51. }
    \label{mean_var}
  \end{center}
\end{figure}

The primary reason for the requirement of accurate experimental 
data is the low sensitivity of the kaon yields to the 
domain volume. At temperatures in the vicinity of 170MeV a large 
number of kaons ($\sim 50$) are produced. This causes the 
kaon yields to approach the grand canonical limit, in other words 
the kaons yields do not experience sufficient canonical suppression. 
A similar trend has recently been observed 
in a partonic scenario at higher temperature \cite{pra03}, 
where very small volumes are 
required to observe canonical suppression.  
Na\"{\i}vely, one would expect multiple strange particles to 
be subject to greater canonical strangeness conservation. 
We present the probability distributions for particles of 
strangeness -2 (\ie, $\Xi$ particles) 
and strangeness -3 (\ie, $\Omega$ particles).
These are plotted in the left panels of Figs. (\ref{probc},\ref{probw}).
We note that, as expected, the difference 
between the distributions 
is more pronounced for the $\Xi$ particles, 
and even more for the $\Omega$ particles, 
as the number of domains is increased.
The $\Xi$ distributions are peaked 
near 1 to 3 particles. The mean number is 
about 2 $\Xi$ particles. The 
$\Omega$ distributions are peaked at zero and the mean 
number is a fraction. We note 
that there is an order of magnitude difference 
between the probability to 
obtain 10 $\Xi$ particles from one domain compared  to 
that from 51 domains. The 
same is true for the probability to obtain 5 
$\Omega$ particles. The minimum probabilities in this region 
are $\sim 10^{-6}$ and hence,
are in principle measurable at RHIC. It should 
be pointed out that, the effect 
of canonical strangeness conservation, for cases 
where we have greater than 20 partitions (or
domain volumes smaller than $20 \, V_{nuc}$),
may be deduced by an accurate determination of the 
mean numbers of kaon, $\Xi$ and 
$\Omega$ particles (see Fig. (\ref{mean_var})). 
From Fig. (\ref{mean_var}) we note that 
a determination of the mean number of $\Omega$ 
particles with an accuracy of 5\%  will allow
one to estimate the domain volume up to a volume 
of $V=20 \, V_{nuc}$ (this is for the case of
full freeze-out volume $V_f = 400 \, V_{nuc}$ and a $T=170$ MeV ). 
For domain volumes greater 
than this, one will need to determine the mean numbers with 
very high accuracy (errors $\sim 1$ \%). Alternatively, one 
may determine the 
probability to obtain 5 $\Omega$ particles with an 
accuracy of 20\%, or the probability to obtain 
10 $\Omega$ particles with an accuracy of 50\%.
Similar error bars may also be deduced for the 
cascades. Which observable is better 
suited to the determination of the domain size, 
now becomes a question of experiment. The behaviour of experimental data 
on the mean numbers of strange particles 
\cite{hoh03} (see also the canonical model analysis in Ref. \cite{ker02}) 
indicates that the number of partitions, in central collisions, is in the 
range from 1 to 20 \ie, the region in Fig. (\ref{mean_var}) 
where the mean values reach a plateau. As a result of this, 
determination of the amount of partitioning of the system 
(\ie, resolving the domain size between $V=20 \, V_{nuc}$ to 
domain size of the order of the freeze-out volume $V=V_f$) will 
require very accurate measurements of the means or the probability 
distributions.

In the above discussion, we have demonstrated that 
the ``edges'' of the probability 
distributions for $\Xi$ particles and $\Omega$ particles 
are sensitive to the partitioning of the 
system (we also pointed out that 
the mean yields also shift by 
about 20\% as we go from 1 to 51 partitions). 
This variation in the probability 
distributions is an indication that these 
particles are experiencing canonical suppression. Recall 
the situation of pion production which is insensitive to 
the partitioning of the system, the pion 
yields are calculated in the grand canonical ensemble. The 
probability distributions from such an ensemble are Poissonian. 
A more quantitative measure of the departure of the $\Xi$ and 
$\Omega$ production from the grand canonical limit may be obtained  
from the ratio of the probability 
distributions to the corresponding 
Poisson distributions.  
Using the mean $\lc n \rc$ we may 
construct a Poisson distribution to obtain 
$n$ $\Xi$ particles from $p$ domains as, 
 
\[
P^n_p = \frac{\lc n \rc^n}{n!}{e^{-\lc n \rc}}.
\]

\nt The ratio of the actual, calculated, probability distribution 
to the Poisson probability distribution 
is plotted in the right panel of Fig. (\ref{probc}) for varying number of 
domains as indicated. 
%The actual calculated 
%probabilities are also plotted at the bottom left corner for scale. 
We note that 
the probability distributions closely resemble a 
Poisson distribution up to the 
probability to obtain about eight $\Xi$ particles. 
Beyond this the distributions begin to 
deviate from a Poisson distribution. 
The same ratio is also plotted for the 
$\Omega$ particles in Fig. (\ref{probw}). 
The $\Omega$ distribution remains 
Poissonian up to five $\Omega$ particles. 

An interesting pattern that may be noted 
both for the $\Xi$ and $\Omega$ distributions is 
that the probability distributions approach 
the Poisson distribution as the 
number of domains is increased compared to 
the number of particles (see the right 
panels of Figs. (\ref{probc},\ref{probw})). 
This effect is much more pronounced in the 
case 
of the $\Omega$ distributions. This may be 
understood as follows: 
in the case of a single, small domain, the 
$\Omega$ probability distribution 
achieves an almost binary configuration. 
The probability $b$
to produce zero $\Omega$ particles is large 
(\ie, $b \ra 1$), followed by a small 
probability $a$ to produce one $\Omega$ 
particle (\ie, $a \ra 0$). The probability to produce multiple 
$\Omega$ particles is very small and hence 
may be ignored (\ie, $a+b = 1$). If there
are $p$ domains, the probability to 
obtain $n$ $\Omega$ particles is given by the 
simple binary distribution,

\bea
P_p^n = \frac{p!}{n! (p-n)!} a^n b^{p-n}. 
\eea

\nt As $p$ becomes much larger than $n$, 
the binary distribution turns into the 
Poisson distribution. This is why the 
ratio of probabilities in 
Figs. (\ref{probc},\ref{probw}) (right panels) 
begins to deviate from 1 as we go to 
higher and higher numbers of particles. 
This corresponds to increasing $n$. 
Increasing the number of domains (keeping 
the volume of each domain constant)
corresponds to increasing $p$. If instead, 
the total volume is held fixed and the 
number of domains is increased, then aside 
from the increase in $p$, the production 
of $\Omega$ particles from each domain approaches 
the binary limit. As a result of these 
two effects, increasing the number of domains 
drives the probability distribution 
function towards the Poisson value. Similar 
effects in moderation may be ascribed 
to the $\Xi$ probability distribution 
functions. Thus, when the number of domains is 
large compared to the number 
of particles produced, one obtains 
a Poisson like probability.

As pointed out in the previous paragraph, 
$\Omega$ production from a single, small 
domain has a binary configuration, 
\ie, a very narrow distribution 
($P(n) \neq 0 $ only for $n=0,\,1$ ).  
The distribution from many such 
domains (see left panel of Fig. (\ref{probc})), is 
necessarily broader due to fluctuations. 
However, even for the case of no partitioning,
the probability to produce two $\Omega$ 
particles from the full freeze-out volume 
is suppressed by a factor of three 
compared to that for a single $\Omega$, 
\ie, the distributions are still rather narrow 
(see left panel of Fig. (\ref{probw})). 
An amusing consequence of this is that 
the variance of the $\Omega$ 
distribution is equal to its mean.
The variance is defined as 

\[
\sigma = \lc n^2 \rc - \lc n \rc^2 = \sum_n  n^2 P(n) - 
\left(\sum_n n P(n)\right)^2 ,
\]
 
\begin{figure}[htb]
\resizebox{3in}{2.75in}{\includegraphics[0in,1in][6.5in,10in]{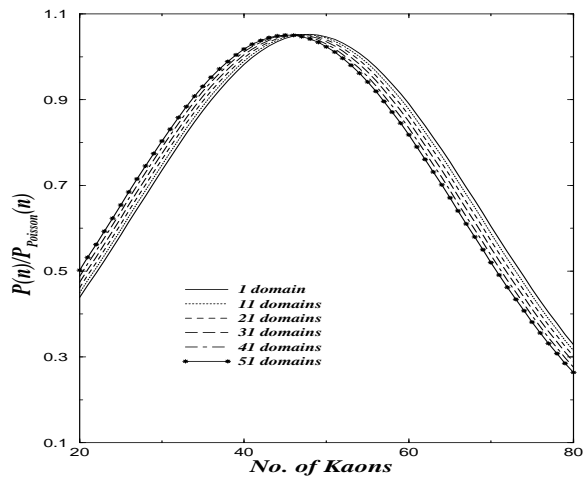}} 
    \caption{ Ratio of the calculated probability distribution 
    of the Kaons to the Poisson distribution. 
    Same as right panel of Fig. (\ref{probc}).}
    \label{probk}  
\end{figure}

\nt
where $P(n)$ is the probability to 
obtain $n$ $\Omega$ particles. 
This equality would indicate 
that the $\Omega$ yields are 
tending towards the grand canonical 
limit, whereas the situation is 
quite the opposite. 
%Finally, we point out that a simple 
%measure of the mean and the variance of the 
%distributions will not allow 
%us to discern the amount of suppression of strange 
%yields due to canonical strangeness conservation.
This is illustrated in Fig. (\ref{mean_var}), 
here we plot the mean and variance of the
the Kaons, $\Xi$ and $\Omega$ particles. 
For a grand canonical ensemble (no suppression), 
the distribution is Poissonian. 
As a result, the variance equals the mean. 
As we note from Fig. (\ref{mean_var}), the 
$\Omega$ particles seem to exhibit a 
Poissonian variance, whereas the $\Xi$ particles deviate slightly 
from this limit. The Kaons display a 
noticeably different variance from the 
Poissonian value. As pointed out previously, 
the reason behind this rather contradictory result is the 
fact that the $\Xi$ and $\Omega$ 
distributions descend rapidly from their peak 
values (see left panels of Figs. (\ref{probc},\ref{probw})). 
The decent is much more mild 
for the kaons (see left 
panel of Fig. (\ref{probfig})). 
All three types of 
particle distributions show a Poissonian 
behavior near the  ``peak'' of the distribution, 
deviating from such behavior as one 
proceeds to the ``edges'' of the distribution. 
The Kaon distribution resembles a Poisson distribution 
over a much wider range than the $\Xi$ or $\Omega$ 
distributions (see Fig. (\ref{probk})).
While the  $\Xi$ and $\Omega$ 
distributions predominantly sample the region in the 
vicinity of the ``peak'', the variance of the 
kaons receives support from 
a wider region. This causes the $\Xi$ and $\Omega$ 
variance to approach that 
of a Poisson distribution, while the kaon variance 
falls below this limit. 
Hence, the appearance 
of a Poisson like variance is 
not indicative of a lack 
of canonical suppression, and vice-versa.

A study of Fig. (\ref{mean_var}) indicates that the 
Kaon variance does not tend towards its mean, even as 
the volume is increased. This may be understood by a 
simple analytical argument. Imagine a large freeze-out 
volume ($V_f > 400 \, V_{nuc}$) at a sufficiently high 
temperature ($T \sim 170 $ MeV). If the only particles 
carrying a conserved strangeness quantum number are 
kaons and anti-kaons, one would 
expect a large number of these 
to be produced ( $\lc N_K \rc = \lc N_{\bar{K}} 
\rc \sim 100 $). In fact, the mean number of kaons $\lc N_K \rc$, 
anti-kaons $\lc N_{\bar{K}} \rc$ and the 
total, mean, number of strangeness carrying particles $\lc N \rc$ 
can now be estimated by a grand canonical ensemble.
Strangeness conservation, however, is imposed by 
the following relation

\bea
N_K + N_{\bar{K}} &=& N \nn \\
N_K - N_{\bar{K}} &=& 0.   \label{vk1}
\eea 

\nt
We may square the above equations, take the mean and sum them to obtain, 

\bea
2 \lc N_K^2 \rc + 2 \lc N_{\bar{K}} ^2 \rc = \lc N^2 \rc.  \label{vk2}
\eea

\nt We may also take the mean and then square and sum to obtain,
   
\bea
2 \lc N_K \rc^2 + 2 \lc N_{\bar{K}}  \rc^2 = \lc N \rc^2 .  \label{vk3}
\eea

\nt We may now subtract Eq. (\ref{vk3}) 
from Eq. (\ref{vk2}) to obtain the 
variance. As the total number of particles 
carrying strangeness, $N$, is not a conserved 
quantity, its distribution may be 
estimated accurately by a grand canonical 
ensemble (\ie, a Poisson distribution).
Hence, the variance of $N$ is equal to its mean, \ie,

\bea
\sigma_N = \lc N \rc = \lc N_K \rc + \lc N_{\bar{K}} \rc.  \label{vk4}
\eea

\nt From the second line of Eq. (\ref{vk1}) we obtain

\bea
\lc N_K \rc &=& \lc N_{\bar{K}} \rc  \nn \\
\sigma_{N_K} &=& \sigma_{N_{\bar{K}}}. \label{vk5}
\eea

\nt
From Eqs. (\ref{vk2}-\ref{vk5}) we obtain,

\bea
\sigma_{N_K} = \frac{N_K}{2}.
\eea

\nt
Thus, the kaon variance is equal to half of 
that of a Poisson distribution, even for a large system. 
%Hence, the distribution will not 
%reach the Poisson limit even for large systems. 
The presence of particles other than 
kaons in the partition sums calculated 
previously allow for more fluctuation of the 
conserved strangeness among different 
strangeness carriers. This increases the variance 
from $\frac{N_K}{2}$ to the value 
plotted in Fig. (\ref{mean_var}).

It should be pointed out that measurements of the variance of the 
$K/\pi$ ratio at the SPS indicate a Poisson value, at 
forward rapidities \cite{afa01}(central region in fixed target experiments). 
This has been shown to be consistent with a Poisson variance 
for the kaon number \cite{jeo99}.
This is in contradistinction with Fig. (\ref{mean_var}). This 
indicates that there is a contamination
of the kaon number from neighbouring rapidity intervals. 
This contamination is essentially random from event to event 
and thus leads to an increased fluctuation of 
the kaon numbers. This will cause the variance to increase 
and lead it to becoming Poissonian. If instead, it were 
possible to measure the kaon yields over a larger 
acceptance, the percentage of random fluctuation in the 
kaon variance would reduce and one should begin to approach the 
value in Fig. (\ref{mean_var}). This serves as an important 
test to ensure the observation of strangeness production from a closed 
volume; \ie, a region whose strange yields are not influenced (event by event) 
by strange yields from neighbouring regions.

%%%%%%%%%%%%%%%%%%%%%%%%%%%%%%%%%%%%%%%%%%%%%%%%%%%%%%%
%%%%%%%%%%%%%%%%%%%%%%%%%%%%%%%%%%%%%%%%%%%%%%%%%%%%%%%
%%%%%%%%%%%%%%%%%%%%%%%%%%%%%%%%%%%%%%%%%%%%%%%%%%%%%%%

\section{Summary and Conclusions}

%%%%%%%%%%%%%%%%%%%%%%%%%%%%%%%%%%%%%%%%%%%%%%%%%%%%%%%%
%%%%%%%%%%%%%%%%%%%%%%%%%%%%%%%%%%%%%%%%%%%%%%%%%%%%%%%%
%%%%%%%%%%%%%%%%%%%%%%%%%%%%%%%%%%%%%%%%%%%%%%%%%%%%%%%%
 
The principal question addressed in this article has been the 
degree of thermalization achieved in heavy-ion collisions, and 
whether this question may be answered within the framework of statistical
models (\ie, without invoking hard or penetrating probes). We 
began with the understanding that complete equilibrium and 
simple superposition of proton-proton collisions are extreme possibilities.
We quantified the degree of thermalization as the size of the 
domain within which thermalization is achieved, independent of 
many other such domains which may be formed. 
The domain was defined as the region over which chemical 
equilibrium (in particular strangeness chemical equilibrium) was achieved. 
Given the fact that 
chemical equilibrium is achieved at a slower rate than kinetic equilibrium,
estimating the size of such a domain would 
set a lower bound on the size of the kinetically equilibrated region.

We further restricted the definition of the domain as the 
volume within which strangeness in strictly 
conserved. A more complete 
calculation with baryon number also strictly 
conserved may also be performed; 
this has been left for a future effort \cite{maj03}. 
Mean baryon number is set by 
means of a chemical potential. 
In this 
preliminary effort, a study of 
the effect of varying chemical potential
was not performed; $\mu$ was set to 
zero throughout. The full freeze-out 
volume was postulated to consist 
of many such domains. Throughout the 
calculation, the full freeze-out
 volume $V_f$ was set to $400 \, V_{nucleon}$ and 
the temperature to $170$ MeV. 
These chosen values of the thermodynamic 
parameters $T, \, V_f, \, \mu$ 
are approximately in agreement with collisions at RHIC. As a result, 
the domain size $V$ became $V_f/p$, where 
$p$ is the number of domains. The estimation of the domain size was  
performed by noting the behavior of the total 
probability distributions of various 
strange particles as a function of $p$.
These results are essentially 
contained in Figs. (\ref{probfig}-\ref{mean_var}).
These probability distributions are, 
in principle, measurable in heavy-ion collisions.

In the calculation of all the above quantities 
(\eg, mean, variance, probability distributions \etc), 
we have made express use of the 
recursion relation technique \cite{cha95,das98,das99}, 
grafted from intermediate 
energy heavy-ion physics. This technique greatly 
simplifies numerical calculations in the 
canonical ensemble with exact conservation conditions. 
%It elegantly replaces the previous complicated techniques 
%involving series of products of Bessel functions.
Two different sets of 
recursion relations have been devised: 
the first one calculates the partition functions, and 
its derivatives in a given domain 
(Eqs. (\ref{recur1},\ref{recur2},\ref{recur3})), while 
the second convolutes the 
probability distributions from different domains 
(Eq. (\ref{lastrecur})). 
Without the use of recursion 
relation techniques, such 
probability distribution 
calculations may become prohibitively difficult.

It is clear from the behavior of the 
mean numbers of Kaons, $\Xi$ and 
$\Omega$ particles 
in Fig. (\ref{mean_var}) and the 
numbers seen in experiments \cite{hoh03}, that 
domain sizes of the order of a 
nucleon volume are eliminated. The question 
reduces to resolving the domain 
size between $V=20 \, V_{nuc}$ ($p=20$) to 
domain size of the order of the 
freeze-out volume $V=V_f$. A definite result for 
$V$ will involve an accurate 
determination of the probability distributions of the 
Kaons, $\Xi$ and $\Omega$ particles, 
and comparison with Figs. (\ref{probfig}-\ref{probw}). 
Alternatively, one may determine 
the yields with much greater accuracy than 
presently imposed in heavy ion 
collisions and compare with Fig. (\ref{mean_var}).
Such a measurement will set a 
lower bound on the size of the region over which 
thermal equilibrium has been 
established. We also pointed out that measurements 
of two particle correlations, such as the variance, 
may lead to incorrect conclusions regarding the amount of 
canonical suppression faced by different strangeness carriers.  

As a corollary, we note that partitioning the 
system introduces a new parameter $p$, 
which essentially changes the yields of 
strange particles keeping the pion yields 
fixed. Hence this provides a natural formalism 
to understand the difference between the 
$K/\pi$ ratio in central collisions of small 
systems compared to non-central collisions 
of large nuclei, with the same $N_{part}$ 
in both cases (see \eg, Ref. \cite{kam03}). Fitting the experimental 
points involves introducing various refinements 
(\eg, Baryon number conservation, temperature and 
volume fluctuations among domains \etc) to 
the rather simple model presented above. 
These and other considerations will be 
addressed in a future effort \cite{maj03}.

%%%%%%%%%%%%%%%%%%%%%%%%%%%%%%%%%%%%%%%%%%%%%%%%%%%%%%%
%%%%%%%%%%%%%%%%%%%%%%%%%%%%%%%%%%%%%%%%%%%%%%%%%%%%%%%
%%%%%%%%%%%%%%%%%%%%%%%%%%%%%%%%%%%%%%%%%%%%%%%%%%%%%%%

\section{Acknowledgment}

%%%%%%%%%%%%%%%%%%%%%%%%%%%%%%%%%%%%%%%%%%%%%%%%%%%%%%%%
%%%%%%%%%%%%%%%%%%%%%%%%%%%%%%%%%%%%%%%%%%%%%%%%%%%%%%%%
%%%%%%%%%%%%%%%%%%%%%%%%%%%%%%%%%%%%%%%%%%%%%%%%%%%%%%%%     

The authors wish to thank X. N. Wang for helpful discussions. 
This work was supported in part by the Natural
Sciences and Engineering Research Council of Canada and by the Director, 
Office of Science, Office of High Energy and Nuclear Physics, 
Division of Nuclear Physics, and by the Office of Basic Energy
Sciences, Division of Nuclear Sciences, of the U.S. Department of Energy 
under Contract No. DE-AC03-76SF00098.

\end{document}